\documentclass[fleqn,usenatbib]{mnras}

\usepackage[T1]{fontenc}
\usepackage{ae,aecompl}

\usepackage{graphicx}	%
\usepackage{amsmath}	%
\usepackage{amssymb}	%
\usepackage{color}
\usepackage[dvipsnames]{xcolor}
\usepackage{booktabs,threeparttable}
\usepackage{soul}
\usepackage{siunitx}

\usepackage{tikz}
\usetikzlibrary{shapes,arrows}

\usepackage{colortbl}
\usepackage{physics}
\usepackage{ulem}

\def\apj{ApJ}%
\def\apjl{ApJ}%
\def\apjs{ApJS}%
\def\ao{Appl.~Opt.}%
\def\aap{A\&A}%
\def\aaps{A\&AS}%
\def\mnras{MNRAS}%
\def\prd{Phys.~Rev.~D}%
\def\prl{Phys.~Rev.~Lett.}%
\def\pasj{PASJ}%
\def\jcap{JCAP}%
\newcommand{\bea}{\begin{eqnarray}}
\newcommand{\be}{\begin{equation}}
\newcommand{\ben}{\begin{enumerate}}
\newcommand{\bi}{\begin{itemize}}
\newcommand{\eea}{\end{eqnarray}}
\newcommand{\ee}{\end{equation}}
\newcommand{\ei}{\end{itemize}}
\newcommand{\een}{\end{enumerate}}

\newcommand{\om}{\Omega_\mr m}
\newcommand{\omb}{\Omega_\mr b}

\newcommand{\ns}{n_{\rm s}}
\newcommand{\w}{w_0}
\newcommand{\wa}{w_{a}}

\newcommand{\mr}{\mathrm}

\usepackage{xspace}

\renewcommand{\P}{\mathbf P}

\newcommand{\bayesfast}{\textsc{bayesfast}\xspace}
\newcommand{\polychord}{\textsc{polychord}\xspace}
\newcommand{\camb}{\textsc{CAMB}\xspace}
\newcommand{\cosmosis}{\textsc{CosmoSIS}\xspace}
\newcommand{\cosmolike}{\textsc{CosmoLike}\xspace}
\newcommand{\hmcode}{\textsc{HMcode2020}\xspace}
\newcommand{\halofit}{\textsc{halofit}\xspace}

\definecolor{azure}{rgb}{0.0, 0.5, 1.0}
\definecolor{darkgreen}{cmyk}{0.85,0.2,1.00,0.2}

\usepackage{siunitx}
\DeclareSIUnit \h {\ensuremath{\mathit{h}}}
\DeclareSIUnit \parsec {pc}
\DeclareSIUnit \deg {deg}

\usepackage{cleveref}

\usepackage{newtxtext,newtxmath}

\title[Cosmology with Roman - Synergies with CMB lensing]{Cosmology with the Roman Space Telescope - Synergies with CMB lensing}

\author[Wenzl et al.]{Lukas Wenzl$^{1}$\thanks{E-mail:ljw232@cornell.edu}, Cyrille Doux$^2$, Chen Heinrich$^3$, Rachel Bean$^1$, Bhuvnesh Jain$^2$, \newauthor Olivier Dor\'e$^{4,3}$, Tim Eifler$^5$ and Xiao Fang$^{5,6}$
\\
$^{1}$ Department of Astronomy, Cornell University, Ithaca, NY, 14853, USA \\
$^{2}$ Department of Physics and Astronomy, University of Pennsylvania, Philadelphia, PA 19104, USA \\
$^{3}$ California Institute of Technology, Pasadena, CA 91125, USA \\
$^{4}$ Jet Propulsion Laboratory, California Institute of Technology, Pasadena, CA 91109, USA \\
$^{5}$ Department of Astronomy/Steward Observatory, University of Arizona, 933 North Cherry Avenue, Tucson, AZ 85721-0065, USA \\
$^{6}$ Berkeley Center for Cosmological Physics, UC Berkeley, CA 94720, USA \\
}

\date{Accepted 2022 March 17. Received 2022 March 15; in original form 2021 December 16}

\pubyear{2022}

\begin{document}
\label{firstpage}
\pagerange{\pageref{firstpage}--\pageref{lastpage}}
\maketitle

\begin{abstract} 
We explore synergies between the Nancy Grace Roman Space Telescope and CMB lensing data to constrain dark energy and modified gravity scenarios. 
A simulated likelihood analysis of the galaxy clustering and weak lensing data from the Roman Space Telescope High Latitude Survey combined with CMB lensing data from the Simons Observatory is undertaken, marginalizing over important astrophysical effects and calibration uncertainties. Included in the modeling are the effects of baryons on small-scale clustering, scale-dependent growth suppression by neutrinos, as well as uncertainties in the galaxy clustering biases, in the intrinsic alignment contributions to the lensing signal, in the redshift distributions, and in the galaxy shape calibration. 
The addition of CMB lensing roughly doubles the dark energy figure-of-merit from Roman photometric survey data alone, varying from a factor of 1.7 to 2.4 improvement depending on the particular Roman survey configuration. Alternatively,  the inclusion of CMB lensing information can compensate for uncertainties in the Roman galaxy shape calibration if it falls below the design goals.
Furthermore, we report the first forecast of Roman constraints on a model-independent structure growth, parameterized by $\sigma_8 (z)$, and on the Hu-Sawicki $f(R)$ gravity as well as an improved forecast
of the phenomenological $(\Sigma_0,\mu_0)$ model. We find that CMB lensing plays a crucial role in constraining $\sigma_8(z)$ at $z>2$, with percent-level constraints forecasted out to $z=4$.
CMB lensing information does not improve constraints on the $f(R)$ model substantially. 
It does, however, increase the $(\Sigma_0,\mu_0)$ figure-of-merit by a factor of about $1.5$.

\end{abstract}


\section{Introduction}
\label{sec:intro}

The physical explanation of the observed late-time acceleration of the universe, so-called dark energy, remains unresolved. Current observational constraints allow for a wealth of gravitational models to fit the expansion history. Many of these models predict deviations in the growth of large-scale structure, galaxies and clusters of galaxies, from that predicted by the current cosmological concordance model called $\Lambda$CDM \citep{Zhao2009}.
The next generation of large-scale structure (LSS) and cosmic microwave background (CMB) observations will be able to measure the growth of structure through a broad range of tracers of the matter distribution and underlying gravitational potential. 
Using data from multiple probes 
in tandem could help find distinctive signatures in cosmic evolution and structure formation which reveal key information about properties of gravity on cosmic scales and the nature of matter in the universe.

Weak lensing and clustering of galaxies are two LSS probes currently being investigated by multiple, ongoing so-called Stage-III programs. These include the
Dark Energy Survey \citep[DES\footnote{\url{https://www.darkenergysurvey.org/}},][]{DESC2005,Krause2017_DESY1,Troxel2018,Abbott2018,Abbott2019,DESCollaboration2021}, the Kilo-Degree Survey \citep[KiDS\footnote{\url{http://www.astro-wise.org/projects/KIDS/}},][]{Kuijken2019,Giblin2020,Heymans2020} and the Hyper Suprime-Cam Subaru Strategic Program \citep[HSC\footnote{\url{http://www.naoj.org/Projects/HSC/HSCProject.html}},][]{Aihara2018,Hamana2020}.
Going forward, multiple planned Stage-IV programs with complementary goals will significantly increase the amount of data available and further improve cosmological constraints, in particular on dark energy. These include the Dark Energy Spectroscopic Instrument \citep[DESI\footnote{\url{https://www.desi.lbl.gov/}},][]{DESICollaboration2016}, the Prime Focus Spectrograph \citep[PFS\footnote{\url{https://pfs.ipmu.jp/}},][]{Takada2014}, the Vera C. Rubin Observatory Legacy Survey of Space and Time \citep[LSST\footnote{\url{https://www.lsst.org/}},][]{Ivezic2019}, Euclid\footnote{\url{https://sci.esa.int/web/euclid}} \citep{Laureijs2011}, the Spectro-Photometer for the History of the Universe, Epoch of Reionization, and Ices Explorer \citep[SPHEREx\footnote{\url{http://spherex.caltech.edu/}},][]{Dore2014} and the Nancy Grace Roman Space Telescope
\citep[Roman\footnote{\url{https://roman.gsfc.nasa.gov/}},][]{Spergel2015}.

\citet{Eifler2020_LSST,Eifler2020_HLS} presented a forecast for the combined analysis of Roman Space Telescope 
galaxy clustering and weak lensing data using a reference design for the High Latitude Survey combined with overlapping data from LSST in visual frequency bands. This included trade-off considerations between a smaller area but deeper survey configuration and shallower observations over a wider area, with an additional wide survey proposed as a suggestion for Roman's possible extended mission period. We use the scenarios in that work as the basis for our modeling of the Roman photometric survey and explore synergies between this survey and CMB lensing data in the context of constraints on models that extend beyond $\Lambda$CDM. %
The next generation of surveys will have increased sample sizes, redshift coverage, and resolution. To translate this increased precision into improved cosmological constraints we need to expand our modeling capabilities to ensure astrophysical effects are accurately modeled and accounted for. %

The lensing of the CMB offers an additional tracer of the gravitational potential which can be reconstructed from CMB temperature and polarization maps. 
Over the last decade a wealth of CMB experiments, including ESA's Planck satellite\footnote{\url{https://www.esa.int/Planck}} \citep{PlanckCollaborationXVII2013,PlanckCollaborationXV2015,PlanckCollaborationVIII2018}, the Atacama Cosmology Telescope \citep[ACT\footnote{\url{https://act.princeton.edu/}},][]{Das2011,Sherwin2017}, the South Pole Telescope \citep[SPT\footnote{\url{https://pole.uchicago.edu}},][]{Story2015,Wu2019}, BICEPT2\footnote{\url{https://lweb.cfa.harvard.edu/CMB/bicep2/}} \citep{BICEP2Collaboration2016} and POLARBEAR\footnote{\url{https://bolo.berkeley.edu/polarbear/}} \citep{Ade2014_polarbear,POLARBEARCollaboration2017}, have measured the lensing of the CMB imparted by the LSS, at high signal-to-noise levels. 
The next generation of CMB experiments such as the Simons Observatory \citep[SO\footnote{\url{https:/simonsobservatory.org/}},][]{SimonsObservatory2019} and CMB-S4\footnote{\url{https://cmb-s4.org/}} \citep{Abazajian2016} will produce significantly more detailed CMB lensing maps of large fractions of the sky.

A range of potential synergies between galaxy clustering and weak lensing as well as CMB lensing has motivated combined analyses of the data. The cross-correlation between CMB lensing and weak lensing was first measured in \citet{Hand2015}. \citet{Singh2019,Singh2020} combined SDSS and Planck data. The DES year~1 data has been analyzed together with CMB lensing data from SPT and Planck \citep{Baxter2019,Omori2019}. They found reduced systematics errors from galaxy weak lensing calibration.  KIDS-1000 data has been combined with CMB lensing data from ACT and Planck \citep{Troester2021,Robertson2021,Sgier2021}.

For Stage-IV programs using the data in tandem is projected to break degeneracies between cosmological parameters and astrophysical systematic uncertainties. 
\citet{Vallinotto2012} argued cross-correlating weak lensing with CMB lensing can help with calibration of the shape detection especially at high redshift, focusing their discussion on LSST. Synergies between LSST and SO were further investigated in \citet{Schaan2017,Fang2021} and synergies between Euclid and CMB probes were investigated in \citet{EuclidCollaboration2021_CMB}.

In this work, we specifically focus on Roman galaxy clustering and weak lensing data combined with CMB lensing data from SO. Throughout this work we will refer to the Roman only tracers as 3x2pt, indicating the three types of auto- and cross-correlations possible when analyzing galaxy clustering and weak lensing. When adding the CMB lensing auto- and cross-correlations we will use the term 6x2pt.
For our baseline model, we are going to study the prospective constraints on a dynamical dark energy, parametrized with $\w \wa$. To produce a realistic forecast we carefully model the expected dominant astrophysical effects. 
This includes the suppression of growth at small scales by massive neutrinos which is expected to become a measurable effect with next-generation experiments \citep{SimonsObservatory2019}. It also includes the effects on baryonic interactions on small scales \citep{Mead2021} and intrinsic alignment (IA) of galaxies for weak lensing analysis which can mimic cosmological signals \citep{Hirata2004}. %
Through conservative cutoffs on the small scales, we avoid regions of significant modeling uncertainties. We also marginalize over shear calibration bias \citep{Massey2013} and redshift estimation uncertainties \citep{Cawthon2018} which are expected to be the most relevant nuisance parameters for cosmological parameter inference \citep{Eifler2020_LSST}. 

Going beyond the baseline forecast, we also explore different scenarios for the prior calibration of the nuisance parameters, as well as a wide survey scenario that trades off depth for area in the context of combining Roman with the SO data.
We also consider extensions to $\Lambda$CDM models that include a model-independent modification to the growth of structure $\sigma_8 (z)$ \citep{SimonsObservatory2019,GarciaGarcia2021}, a phenomenological model for deviations in the gravitational potentials \citep{2021MNRAS.tmp.2957L, DESy1_extendedmodels, 2019PhRvD..99h3512F}, and the Hu-Sawicki $f(R)$ model \citep{Hu2007}. For our inferred parameter constraints we go beyond a Fisher analysis to capture non-gaussianity in parameter degeneracies.

The paper is laid out as follows: In \Cref{sec:formalism} we discuss how we model the observable power spectra. We then discuss how we model the full non-linear covariance matrix for the observables in \Cref{sec:cov}. In \Cref{sec:sampling} we discuss our approach for the parameter inference. Then in \Cref{sec:results} we present the results of our analysis and conclude with a discussion in \Cref{sec:discussion}.

\section{Theory and Modeling} \label{sec:formalism}

In this section, we describe the forecasted observables in detail. We first describe the galaxy clustering and weak lensing dataset from Roman in \Cref{sec:Roman} and the CMB lensing dataset from SO in \Cref{sec:SO}. We then describe our baseline cosmology in \Cref{sec:baselinecosmo}. We lay out the modeling pipeline to calculate the full set of observables based on our set of varied cosmological and nuisance parameters. Finally, we describe the extended cosmological models in \Cref{sec:extendedcosmologicalmodels}.

\subsection{Roman Space Telescope}\label{sec:Roman}

{\renewcommand{\arraystretch}{1.2}%
\begin{table*}
\begin{tabular}{|lll|lllllll|}
\hline
\multicolumn{3}{|c|}{\textbf{Roman Space Telescope}}                                                                   & \multicolumn{7}{c|}{\textbf{Simons Observatory}}                                                                                                               \\ \hline
                                                                & HLS survey            & Wide survey         & \multicolumn{7}{c|}{LAT}                                                                                                                              \\ \hline
Sky area                                                        & \SI{2000}{\square\deg}    & \SI{18000}{\square\deg} & $f_{\rm{sky}}$              & 0.4                &                    &                   &                   &                   &                   \\
Number of sources (weak lensing)                                & $51\, \rm{gal/arcmin}^{2}$   & $43\, \rm{gal/arcmin}^{2}$ & Frequency band (\SI{}{\GHz})        & 27                 & 39                 & 93                & 145               & 225               & 280               \\
Number of lenses (clustering)                                   & $66\, \rm{gal/arcmin}^{2}$   & $50\, \rm{gal/arcmin}^{2}$ & FWHM (arcmin)                    & 7.4                & 5.1                & 2.2               & 1.4               & 1.0               & 0.9               \\
$ \sigma_\epsilon \cdot \sqrt{2} $&  0.37  & 0.37     & White Noise (\SI{}{\micro\kelvin}-arcmin) & 71                 & 36                 & 8.0               & 10                & 22                & 54                \\
Scenarios (see \Cref{tab:varied_params}) & optmistic/conservative & wide    & Full Noise Curve            & \multicolumn{6}{c|}{\href{https://github.com/simonsobs/so_noise_models}{so\_noise\_models v3.0.0}} \\ \hline
\end{tabular}
\caption{Properties of the surveys considered. For the Roman Space Telescope we consider the HLS survey as well as a proposed wide survey. For the Simons Oberservatory we use the baseline scenario from \citet{SimonsObservatory2019}.\label{tab:surveys} }

\end{table*}}

The Roman Space Telescope\footnote{The Roman Space Telescope was formerly called Wide-Field InfraRed Survey Telescope (WFIRST).} is an upcoming NASA flagship mission with a diverse portfolio of science goals around cosmology, exoplanets, and general astrophysics \citep{Spergel2015}. It was selected as the top priority in the 2010 decadal survey and is now in its design and fabrication phase (Phase C).

One of the major parts of the mission is the High Latitude Survey (HLS) with both an imaging and a spectroscopic component \citep{Wang2021}. In the current reference imaging survey, Roman would scan over a sky area of $\SI{2000}{\deg\squared}$ in a total time of 1.9 years to produce a sample of galaxies for clustering and weak lensing studies. In this configuration, Roman is projected to find about $\SI{51}{gal \per arcmin \squared}$ viable for weak lensing analysis and $\SI{66}{gal \per arcmin \squared}$ viable for clustering analysis \citep{Eifler2020_HLS}\footnote{This estimate does assume the use of overlapping optical data from Rubin.}. We will consider an optimistic and a conservative scenario for the priors we place on the nuisance parameters. 

In this paper, we also compare an alternative wide survey \citep[wide,][]{Eifler2020_LSST} which aims to observe the entire LSST footprint in one band to comparable depth. This extends LSST by a redder W band, leading to a Rubin+RomanW dataset for clustering and weak lensing. In this shallower scenario, the survey is projected to result in about $\SI{43}{gal \per arcmin \squared}$ viable for weak lensing analysis and $\SI{59}{gal \per arcmin \squared}$ viable for clustering analysis \citep{Eifler2020_LSST}. The actual HLS survey will be decided shortly before launch and could include such a wide component. Also, as the Roman Space telescope's lifetime will not be limited by any consumables, it is possible to extend the mission beyond its nominal five-year lifespan, which would open up the possibility for an additional large survey. For the wide survey we make similar prior assumptions as for the optimistic HLS scenario, except for a larger redshift uncertainty due to less photometric information.
We summarize the HLS and wide survey configurations in \Cref{tab:surveys}. 

\begin{figure}
 \includegraphics[width=8.5cm]{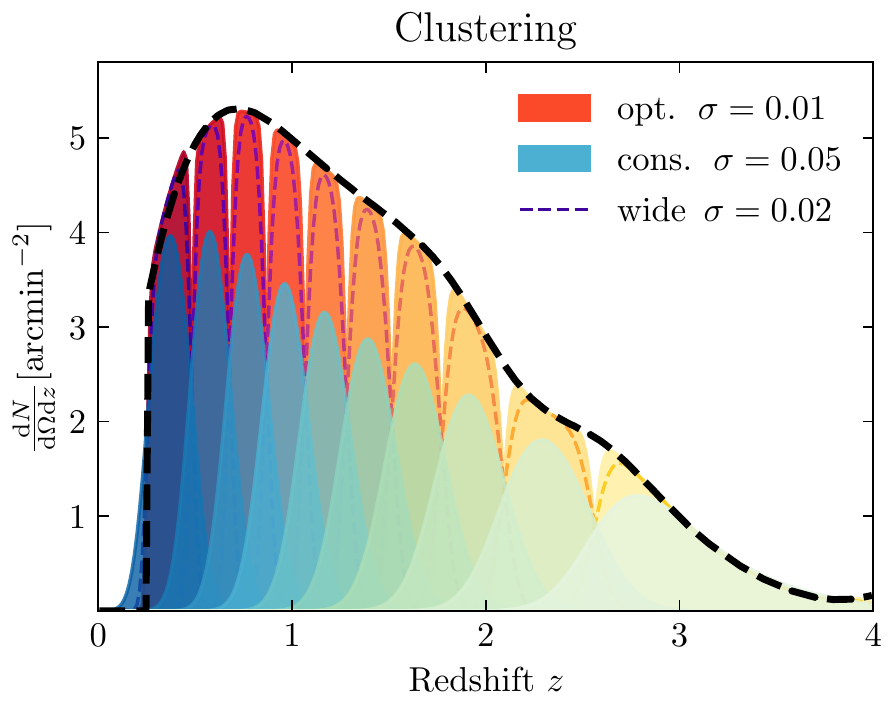}
 \includegraphics[width=8.5cm]{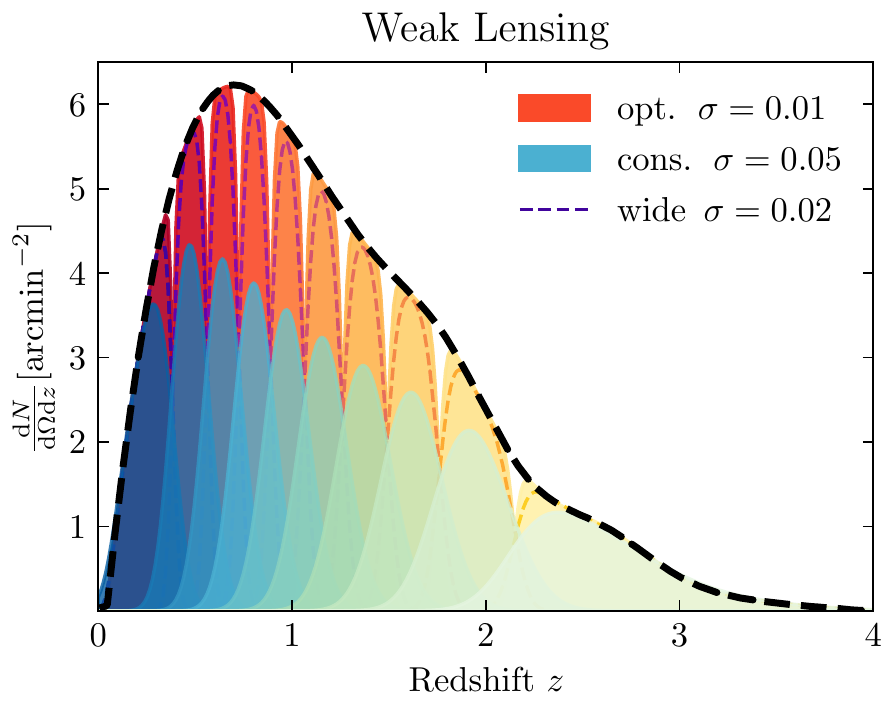}
 \caption{Redshift distributions of the ten tomographic clustering [top] and weak lensing [bottom] bins. Shown are the optimistic case [red] with $\sigma_z = 0.01$, wide case [thin dashed line] with $\sigma_z = 0.02$ and conservative case [blue] with $\sigma_z = 0.05$ relative to the overall distribution [thick black dashed line].\label{figure:bins}}
 \end{figure}

We follow \citet{Eifler2020_LSST} and \citet{Eifler2020_HLS} and the assumption therein for modeling the galaxy redshift distributions. 
In particular we take the same overall redshift distributions $n^{\textrm{lens}} (z)$ and $n^{\textrm{source}} (z)$ for the galaxy clustering and weak lensing samples respectively.
We split each set into 10 bins with equal amounts of objects.
We model photometric redshift uncertainties with a shift $\Delta_{z, X}^i$ for each bin and an overall scatter $\sigma_{z,X} (1+z)$ for both our clustering ($X = \textrm{lens}$) and weak lensing samples ($X = \textrm{source}$). For each we use 10 tomographic bins with equal amounts of objects split along redshift. To model the redshift uncertainties, we convolve each bin with a Gaussian kernel, given by
\begin{equation}
    p^i(z|z', X) \equiv \frac{1}{\sqrt{2\pi} \sigma_{z,X}(1+z)} \textrm{exp}\left(- \frac{(z' - z - \Delta_{z,X}^i)^2}{2 (\sigma_{z,X}(1+z) )^2}\right).
\end{equation}
For given boundaries $z^i_{\textrm{min}}, z^i_{\textrm{max}}$ of a tomographic bin $i$, we compute the number density as
\begin{equation}
    n_i^X(z) \equiv \int_{z^i_{\textrm{min}}}^{z^i_{\textrm{max}}} dz' n^X (z') p^i(z|z', X).
\end{equation}
\Cref{figure:bins} shows the number densities for our fiducial tomographic bins for the different survey scenarios considered. The graph also shows the overall assumed distributions of galaxies as a black dashed line. 

\subsection{Simons Observatory} \label{sec:SO}

The Simons Observatory is an upcoming CMB experiment consisting of one large-aperture telescope (LAT) and three small-aperture telescopes \citep{SimonsObservatory2019}. The LAT will produce a CMB lensing map of 40\% of the sky, including the entire HLS footprint and overlapping with the majority of LSST, and therefore also the proposed wide survey. SO will observe six frequency bands for which we use the baseline noise characteristics given in \citet{SimonsObservatory2019} and listed in \Cref{tab:surveys}. With the assumed minimum variance combination of quadratic estimators, the reconstructed lensing convergence map $\kappa$ is projected to be signal dominated up to about $\ell=200$. We do not include the CMB temperature and polarization power spectra in our analysis.

\subsection{Baseline Cosmological Model}
\label{sec:baselinecosmo}

Our baseline cosmological model comprises eight core cosmological parameters: the fractional density in matter today, $\om$, the dimensionless Hubble constant, $h =H_0 / (\SI{100}{\km \per \second \per \mega \parsec})$, the fractional baryon density today $\omb$, the scalar index of the primordial power spectrum $\ns$, the amplitude of dark matter fluctuations averaged on scales of $\SI{8}{\per\h\mega \parsec}$, $\sigma_8(z=0)$, the neutrino density, $\Omega_\nu h^2$, 
and 
a dark energy model with a time-varying equation of state given by $w(a) = \w + (1-a) \wa $.
We list the full set of parameters for our cosmological model in \Cref{tab:varied_params}. The fiducial values for cosmological parameters are based on \citet{PlanckCollaboration2018}. For each of these parameters, we use a flat prior well outside the level of current constraints. For analyzing data from Roman and SO the effect of neutrinos is expected to be relevant and needs to be accounted for to forecast realistic constraints on the growth of structure. Neutrinos introduce a scale dependence for the growth function, sensitive to the sum of the neutrino masses, as growth is suppressed below the free-streaming scale of the neutrinos. We adopt a model with $N_{\rm eff} = 3.046$ where one eigenstate is massive, and we use the physical energy density ($\Omega_\nu h^2$) as a free parameter. For our fiducial value we use the lower limit from particle experiments of $\sum m_\nu =\SI{0.06}{\eV}$ \citep{Patrignani2016_pdg}.
We assume a flat universe by enforcing $\Omega_\Lambda = 1 - \om$.  
In \Cref{tab:varied_params} we show the prior ranges, list the modified gravity scenarios we are considering in \Cref{sec:extendedcosmologicalmodels} and list all nuisance parameters we marginalize over.

Constraints on the $\w \wa$ parameterization are considered an important benchmark for Stage-IV dark energy programs. We evaluate the constraining power via the so-called figure-of-merit for dark energy (FoM) as defined in \citet{DarkEnergytaskforce2006}: 
\begin{equation}
   \textrm{FoM} \equiv (\textrm{det Cov}(\w, \wa))^{-1}, \label{eq:FoM}
\end{equation}
where $\textrm{Cov}$ is the covariance matrix to our posterior samples. 

For the fiducial values, we use the cosmological constant scenario, i.e. $\w=-1$ and $\wa=0$.  

To build our pipeline, we use \cosmosis\footnote{\url{https://bitbucket.org/joezuntz/cosmosis}} \citep{Zuntz2015}. This modular code for cosmological parameter estimation already includes part of the modeling capabilities we need and has convenient wrappers for parameter inference as we will discuss in \Cref{sec:sampling}.
The following paragraphs show, step by step, how we construct the observable angular power spectra $C (\ell)$ from our cosmological and nuisance parameters for the full 6x2pt data vector.

\newcommand{\greyrule}{\arrayrulecolor{black!30}\hline\arrayrulecolor{black}}
{\renewcommand{\arraystretch}{1.2}%
\begin{table}
\begin{tabular}{@{}ccc@{}}
\hline
\multicolumn{1}{|c|}{Parameter}                                            & Fiducial value        & \multicolumn{1}{|c|}{Flat Prior [range]}                                                                                                                 \\[2pt] \hline 
\multicolumn{3}{c}{\textbf{Baseline cosmology ($\w \wa \textrm{CDM}\nu$)}}                                                                                                                                                             \\[2pt] \hline
\multicolumn{1}{|c}{$\Omega_{\rm m}$}                      & 0.3156                & \multicolumn{1}{c|}{ $[0.1, 0.6]$}                                                                                \\
\multicolumn{1}{|c}{$h$}                           & 0.6727                & \multicolumn{1}{c|}{ $[0.55,0.9]$}                                                                                \\
\multicolumn{1}{|c}{$\Omega_{\rm b}$}                      & 0.04917               & \multicolumn{1}{c|}{ $[0.01, 0.12]$}                                                                              \\
\multicolumn{1}{|c}{$n_{\rm s}$}                           & 0.9645                & \multicolumn{1}{c|}{ $[0.87, 1.07]$}                                                                              \\
\multicolumn{1}{|c}{$\sigma_{\rm 8}$}                      & 0.8310                & \multicolumn{1}{c|}{ $[0.6, 1.1]$}                                                                                \\
\multicolumn{1}{|c}{$\Omega_\nu h^2$}                & $6.155 \cdot 10^{-4}$ & \multicolumn{1}{c|}{ $[0.6, 10] \cdot 10^{-4}$}                                                                     \\ %
\multicolumn{1}{|c}{$\w$}                           & $-1$                    & \multicolumn{1}{c|}{ $[-2.5, 0.5]$}                                                                        \\
\multicolumn{1}{|c}{$\wa$}                           & 0                     & \multicolumn{1}{c|}{ $[-2.5,2.5]$}                                                                            \\[3pt] \hline
\multicolumn{3}{c}{\textbf{Extended cosmologies ($\w, \wa$ fixed)}}                                                                                                                                                       \\[2pt] \hline 
\multicolumn{3}{|c|}{\textbf{$\sigma_8(z)$ scenario ($\sigma_8 $ fixed)}}                                                                                                                                             \\[2pt]
\multicolumn{1}{|c}{$\hat \sigma_8(z_j), z_j = 0,1,2,3,4$}                      & 1                     & \multicolumn{1}{c|}{[0,2]}                                                                                                \\ \hline
\multicolumn{3}{|c|}{\textbf{$\mu_{\rm 0} \Sigma_{\rm 0}$ scenario}}                                                                                                                                             \\[2pt]
\multicolumn{1}{|c}{$\mu_{\rm 0}$}                        & 0                     & \multicolumn{1}{c|}{[-3,3]}                                                                                                \\
\multicolumn{1}{|c}{$\Sigma_{\rm 0}$}                      & 0                     & \multicolumn{1}{c|}{[-3,3]}                                                                                                \\ \hline
\multicolumn{3}{|c|}{\textbf{$ f(R)$ scenario (fR6/fR7)}}                                                                                                                                             \\[2pt]
\multicolumn{1}{|c}{$f_{R0}$}                      &  $10^{-6}/10^{-7}$                     & \multicolumn{1}{c|}{[0,$10^{-4}$]/ [0,$10^{-5}$]}                                                                                             \\[2pt] \hline
\multicolumn{3}{c}{\textbf{Nuisance parameters}}                                                                                                                                                       \\[2pt] \hline 
\multicolumn{3}{|c|}{\textbf{Galaxy Bias}}                                                                                                                                                           \\[2pt]
\multicolumn{1}{|c}{$b_{\rm g}^i, i \in \{1,\dots , 10\}$}            & $1.2+ i \cdot 0.1$    & \multicolumn{1}{c|}{ $[0.8, 3]$}                                                                                  \\ \hline
\multicolumn{3}{|c|}{\textbf{Intrinsic Alignment}}                                                                                                                                                   \\[2pt]
\multicolumn{1}{|c}{$A_{\textrm{IA}}$}                           & $5.95$                & \multicolumn{1}{c|}{ $[0, 10]$}                                                                            \\
\multicolumn{1}{|c}{$\eta_{\textrm{IA}}$}                          & $0.49$                & \multicolumn{1}{c|}{ $[-10,10]$}                                                                                    \\
\multicolumn{1}{|c}{$\eta^{\text{high-z}}_{\textrm{IA}}$}           & $0$                   & \multicolumn{1}{c|}{ $[-1,1]$}                                                                                    \\ \hline
\multicolumn{3}{|c|}{\textbf{Baryonic Effects}}                                                                                                                                                      \\[2pt]
\multicolumn{1}{|c}{$\text{log }T_{\text{AGN}}$}                          & $7.8$                 & \multicolumn{1}{c|}{ $[7.6, 8.3]$   }                                                                                                  \\ \hline
\multicolumn{1}{|c}{}                                            &        & \multicolumn{1}{|c|}{Gaussian Prior ($\sigma$)}                                                                                                                 \\ \hline
\multicolumn{3}{|c|}{\textbf{Shear Calibration (optimistic/ conservative/ wide)}}                                                                                                                           \\[2pt]
\multicolumn{1}{|c}{$m^{i}$}                           & $0$                   & \multicolumn{1}{c|}{\begin{tabular}[c]{@{}l@{}}$(0.002)$/ \\ $(0.01)$/ \\ $(0.002)$\end{tabular}} \\ \hline
\multicolumn{3}{|c|}{\textbf{Photometric Redshifts (optimistic/ conservative/ wide)}}                                                                                                                                     \\[2pt]
\multicolumn{1}{|c}{$\Delta_{z, \textrm{source}}^i$} & $0$                   & \multicolumn{1}{c|}{\begin{tabular}[c]{@{}l@{}}$(0.001)$/\\ $(0.02)$/ \\ $(0.001)$\end{tabular}}                     \\\greyrule
\multicolumn{1}{|c}{$\sigma_{z, \textrm{source}}$}   & \multicolumn{1}{c}{\begin{tabular}[c]{@{}l@{}}$0.01$/ \\ $0.05$/ \\ $0.02$\end{tabular}}                & \multicolumn{1}{c|}{\begin{tabular}[c]{@{}l@{}}$(0.002)$/ \\ $(0.02)$/ \\ $(0.002)$\end{tabular}}                  \\ \greyrule
\multicolumn{1}{|c}{$\Delta_{z, \textrm{lens}}^i$}   & $0$                   & \multicolumn{1}{c|}{\begin{tabular}[c]{@{}l@{}}$(0.001)$/\\ $(0.02)$/ \\ $(0.001)$ \end{tabular}}                       \\ \greyrule
\multicolumn{1}{|c}{$\sigma_{z, \textrm{lens}}$}     & \multicolumn{1}{c}{\begin{tabular}[c]{@{}l@{}}$0.01$/ \\ $0.05$/ \\ $0.02$\end{tabular}}                & \multicolumn{1}{c|}{\begin{tabular}[c]{@{}l@{}}$(0.002)$/ \\ $(0.02)$/ \\ $(0.002)$\end{tabular}}                  \\ \hline
\end{tabular}
\caption{Fiducial values and priors for all varied parameters. Shown is the assumed cosmology with separated sections for the extended cosmologies considered. Additionally all marginalized nuisance parameters are listed. %
\label{tab:varied_params} }
\end{table}}

\subsection{Angular power spectra}\label{sec:angspec}

We model the observable data vector comprised of all types of angular (cross) power spectra $C (\ell)$ based on a given set of cosmological and nuisance parameters. We divide the Roman weak lensing and galaxy clustering samples each into 10 tomographic bins, labeled with $i$ and $j$.

To obtain the observable angular power spectra, we project the matter power spectrum $P_{\delta \delta}$ onto the sky weighted by a function specific to each observable. Angular power spectra of the projected fields are obtained in the Limber approximation by,
\begin{equation}
C_{A B}^{i j}(\ell)=\int \dd{\chi} \frac{W_{A}^{i}(\chi) W_{B}^{j}(\chi)}{\chi^{2}} P_{\delta \delta}\qty(k=\frac{\ell+1/2}{\chi}, z(\chi)),
\label{eq:cl}
\end{equation}
with $A,B \in \{ G, \delta_{\rm g} , \kappa \} $ where $G$ refers to weak lensing, $\delta_{\rm g}$ refers clustering and $\kappa$ refers to CMB lensing. Note that we explicitly distinguish between the cosmological galaxy weak lensing shear $G$ caused by the intervening gravitational potential, and the observed weak lensing shear $\gamma$ which includes the added contribution from IA. %
Here $\chi$ is the comoving distance and $\ell$ designates the multipole. We use 20 $\log$-spaced bins between $\ell=30$ and 3000 for our analysis, for all auto- and cross-power spectra. For weak lensing and CMB lensing, we calculate the power spectra up to $\ell=3000$. For galaxy clustering, we avoid the smallest scales by using a conservative real space cutoff of $R_{\textrm{max}} = 2 \pi / k_{\textrm{max}} = \SI{21}{\per\h\mega\parsec}$, related to $\ell$-space via the Limber approximation $k = \ell /\chi(z)$. For cross-correlations between weak lensing and galaxy clustering we only use bin combinations where the clustering sample is in front of the source sample for weak lensing, i.e. at lower redshift. 

We calculate the linear matter power spectrum using the fitting functions from \citet{Eisenstein1999}. Note that these fitting functions model the scale-dependent suppression of growth from neutrinos correctly; they also neglect baryon acoustic oscillations which is not a problem here since our observables cannot resolve them. To calculate the non-linear corrections to the power spectrum we use the \hmcode\footnote{\url{https://github.com/alexander-mead/HMcode}} from \citet{Mead2021}. 
We also account for baryonic interactions which affect the matter power spectrum on small scales. We partially account for this via scale cuts as described above. However as shown in, for example, \citet{Huang2019}, to use shear measurements up to $\ell = 3000$ in Stage-IV surveys, we additionally need to marginalize over models of the baryonic effects. 
We use the one-parameter halo-model described in \citet{Mead2021}, which uses the active galactic nuclei (AGN) temperature $T_{\textrm{AGN}}$ as a free parameter. We adopt their conservative prior range of $7.6 < \log T_{\textrm{AGN}} < 8.3$. We calculate the nonlinear matter power spectrum $P_{\delta \delta}(k,z)$ up to $k = \SI{10}{\h\per\mega\parsec}$,  %
which is sufficient for our observables, and beyond which we extrapolate by fitting a power law.

Kernel functions $W_A(\chi)$ are specific to each tracer $A$. For galaxy clustering, they depend directly on the redshift distribution of our observed clustering sample for each tomographic bin. For each of the 10 redshift bins defined in \Cref{sec:Roman}, the kernel is given by
\begin{equation}
    W_{\delta_{\rm g} }^{i}(\chi)=b_{\rm g}^i \frac{n^{\text {lens }}_{i}(z(\chi))}{\bar{n}^{\text {lens }}_{i}} \dv{z}{\chi},\label{eq:Wdelta}
\end{equation}
where $\bar n_i^{\textrm{lens}}$ is the average number density of galaxies per bin, and $b_{\rm g}^i$ is the galaxy bias connecting the dark matter density field to the observed galaxy density field.  
We assume that on the scales we consider, the clustering of galaxies 
can be predicted from the underlying matter fluctuations by multiplying with a bias factor $b_{\rm g}^i$ that will be fitted for each bin $i$. We assume a conservative flat prior of the range $[0.8, 3]$ and use fiducial values $b_{\rm g}^i=1.2+ 0.1\ i$, with $i\in [1,10]$. The specific choice of fiducial values here does not have a significant effect on the forecasted constraints. %

The lensing efficiency for a source at comoving distance $\chi_{\rm S}$, which is relevant for galaxy weak lensing and CMB lensing, can be estimated as \citep{Schaan2017},
\begin{equation}
    W_{\kappa}\left(\chi, \chi_{S}\right)=\frac{3}{2}\left(\frac{H_{0}}{c}\right)^{2} \Omega_{\rm m}
    \frac{\chi}{a(\chi)}\left(1-\chi / \chi_{\rm S}\right).
\label{eq:lensing_kernel_fixed_source}
\end{equation}
For CMB lensing, we have a single source, and $\chi_{\rm S}$ corresponds to the comoving distance to the last scattering surface $\chi_*$. For this calculation, we use \camb\footnote{\url{https://camb.info}} \citep{Lewis2011} (version Jan15 included in \cosmosis) in a reduced mode that only calculates the thermal history to derive the lensing kernels $W_{\kappa}(\chi, \chi_*)$. %
For weak lensing, we do not have a single source, but rather need to integrate over our source galaxy distributions to obtain the lensing kernel, given by
\begin{equation}
    W_{G}^i (\chi) = \int_\chi^\infty \dd{\chi'} \frac{n^{\text {source }}_{i}(z(\chi))}{\bar n_i^{\textrm{source}}} \dv{z}{\chi'} W_\kappa (\chi, \chi' ).
\label{eq:lensing_kernel}
\end{equation}

\subsubsection{Intrinsic alignment}\label{sec:IA}

Apart from gravitational shear, the observed ellipticity of galaxies is also influenced by their intrinsic alignment (IA) with the local gravitational potential. The exact IA depends on the complex history of galaxy evolution and will be a major uncertainty for weak lensing constraints from next-generation surveys like Roman \citep{Eifler2020_HLS}. %
Accurately accounting for the IA is an ongoing effort that involves modeling, theory, and improving constraints from observations \citep{Hirata2004,Bridle2007,Krause2016}. 

For our analysis, we will follow the approach laid out in \citet{Krause2016}. We adopt the so-called non-linear alignment (NLA) model \citep{Catelan2001,Hirata2004,Bridle2007} where red elliptical galaxies are assumed to align with the tidal alignment of their host dark matter halos. The shape alignment is then proportional to fluctuations in the large-scale tidal gravity field. While this approach is expected to capture the majority of the IA, it neglects tidal torquing mechanisms \citep{Blazek2019} %
and the sub-dominant alignment of blue spiral galaxies \citep{Samuroff2019}. Furthermore, we neglect non-linear and post-Born effects in CMB lensing cross-correlations which may become relevant at high redshift with SO data \citep{Fabbian2019}. These can mimic the impact of IA on cross-correlations and require additional modeling. We leave an exploration of this to future work.

The auto-correlation of the IA captures the correlation of the tidal field for galaxies that are physically close to each other (II). Additionally, the tidal field also alters the lensing of background sources, leading to significant contributions of the cross-correlation between the IA of the lens and the weak lensing of background sources (IG).  Both effects add to the spatial correlation of cosmic shear (GG). The cross-correlation of cosmic shear with CMB lensing is also affected by an IG type contribution \citep{Larsen2016} and we also include this in our analysis.
In total the observed power spectrum becomes %
\begin{eqnarray}
 C_{\gamma \gamma}^{ij} (\ell) &=& C_{\mathrm{GG}}^{ij} (\ell) + C_{\mathrm{GI}}^{ij} (\ell) +
 C_{\mathrm{IG}}^{ij} (\ell) +C_{\mathrm{II}}^{ij} (\ell),  \\
C_{\gamma \mathrm{B}}^{ij} (\ell) &=& C_{\mathrm{GB}}^{ij} (\ell) + C_{ \mathrm{IB}}^{ij} (\ell),
\end{eqnarray} 
where $B \in \{ \delta_g, \kappa \} $.
We obtain these power spectra by applying the Limber approximation to the matter spectra and multiplying with the fraction of red galaxies $f_{\mathrm{red}}(z) $ and the amplitude $A(z)$ for the IA terms
\begin{eqnarray}
C_{\mathrm{II}}^{i j}(\ell) &=& \int \mathrm{d}\chi \frac{W_{I}^{i}(\chi) W_{I}^{j}(\chi)}{\chi^{2}} f_{\mathrm{red}}^{2}\left( z\right) A^2(z) P_{\delta \delta}, \\
C_{\mathrm{IB}}^{i j}(\ell) &=&\int \mathrm{d}\chi \frac{W_{I}^{i}(\chi) W_{B}^{j}(\chi)}{\chi^{2}} f_{\mathrm{red}}\left( z\right) (-A(z)) P_{\delta \delta}.
\end{eqnarray}
Note that IA is anti-correlated with the other tracers. The IA kernel is given by
\begin{equation}
    W_{I}^{i}(\chi)=\frac{n^{\text {lens }}_{i}(z(\chi))}{\bar{n}^{\text {lens }}_{i}} \dv{z}{\chi}.
\end{equation}
To marginalize $A(z)$ and $f_{\mathrm{red}}$ over luminosity, we use the GAMA luminosity functions from Table~5 of \citet{Loveday2012}, i.e. the parameters $\phi^*_0,M^*,\alpha,Q, P$ ($r$-band) for a Schechter form, given by %
\begin{equation}
\phi(L, z)=\phi^{*}(z)\left(\frac{L}{L^{*}(z)}\right)^{\alpha} \exp \left(-\frac{L}{L^{*}(z)}\right),
\end{equation}
where $\phi^{*}(z)=\phi_{0}^{*} 10^{0.4 z P}$ %
and $L^{*}(z)$ is the luminosity corresponding to
$M^{*}(z)=M^{*}(0.1)-(z-0.1)Q$. For the limiting magnitude of the survey, we assume an equivalent $r$-band magnitude of ${m_{\textrm{lim}} = 25.3}$, as done in \citet{Eifler2020_LSST}. In order to integrate, we also need the luminosity $L(m_{\lim}, z)$ corresponding to
\begin{equation}
M_{\lim }\left(z, m_{\lim }\right)=m_{\lim }-\left(5 \log _{10} \frac{D_{\mathrm{L}}(z)}{\mathrm{Mpc} /h}+25+k(z)\right),
\end{equation}
where $D_{\mathrm{L}}$ is the luminosity distance. The conversion to luminosity is done with $L(m_{\lim}, z) = \bar L \cdot 10^{-0.4 M_{\lim}(m_{\lim},z)}$, with zero-point luminosity $ \bar L = \SI{3.0128e28}{\watt}$ \citep{Mamajek2015}. The same conversion is used for $M^*(z)$ and $M_0(z)$. $k(z)$ is the $k$- and $e$-correction for the $r$-band, taken from \citet{Poggianti1997}. For red galaxies, we use their values for E type galaxies, and for the sample of all galaxies, we use their values for Sa type galaxies.
These corrections only go up to redshift of 3, beyond which we extrapolate linearly.

Given these luminosity functions, our IA amplitude is modeled as
\begin{align}
\begin{split}
    A(z)=\frac{C_{1} \rho_{\mathrm{m}, 0}}{D(z)} &A_{\textrm{IA}}
    \expval{\qty(\frac{L(z)}{L_0(z)})^{\beta_{\textrm{IA}}}}_{\phi_{\mathrm{red}}}\times \\
    &\left(\frac{1+z}{1+z_{0}}\right)^{\eta_{\textrm{IA}}}\begin{cases}
       1 & z < z_1\\
      \left(\frac{1+z}{1+z_1}\right)^{\eta^{\textrm{high-z}}_{\textrm{IA}}} & z \geq z_1
    \end{cases},
\end{split}
\end{align}
following \citet{Joachimi2011}, but with an additional parameter $\eta^{\textrm{high-z}}_{\textrm{IA}}$, as done in \citet{Krause2016}. 
Therefore, the amplitude of IA depends on four parameters in our model: an overall amplitude factor $A_{\textrm{IA}}$, an exponent $\beta_{\textrm{IA}}$ for the luminosity evolution, an uncertainty parameter $\eta_{\textrm{IA}}$ for the redshift evolution, and a second uncertainty parameter $\eta^{\textrm{high-z}}_{\textrm{IA}}$ for the high redshift evolution, because we are extrapolating any priors on IA from current observations.
As fiducial values, we use the best-fit parameters for the MegaZ-LRG + SDSS LRG sample from \citet{Joachimi2011}. 
We also use their $z_0 = 0.3$ and set $z_1 = 0.7$ since their fits extends to about $z \leq 0.7$. For the normalization, we use the fit to the SuperCOSMOS observations \citep{Hirata2004,Bridle2007} giving $C_1 \rho_{\textrm{crit}}= 0.0134$, and $D(z)$ is the growth function. 
We marginalize the luminosity dependence over the luminosity function of red galaxies. This depends on the choice of luminosity function and the limiting magnitude of our survey, ${m_{\textrm{lim}}}$, which sets the lower cutoff of the integral. Finally, $L_0$ is the characteristic luminosity corresponding to absolute $r$-band magnitude $M_0 = -22-Qz$. 

The fraction of red galaxies is given by 
\begin{equation}
    f_{\mathrm{red}}\ (m_{\lim} , z  ) = \frac{\int_{L(m_{\lim} , z)}^\infty \dd{L}    \phi_{\mathrm{red}}(L, z)}{\int_{L(m_{\lim} , z)}^\infty  \dd{L} \phi_{\mathrm{all}}(L, z)  },
\end{equation}
which depends on $m_{\textrm{lim}}$, the limiting magnitude of our survey, and both luminosity functions $\phi_{\rm red}$ and $\phi_{\rm all}$.

For our choice of luminosity function, we observe a strong degeneracy between $\beta_{\textrm{IA}}$ and $A_{\textrm{IA}}$. Our inference scheme struggles to capture the irregular shape of the degeneracy. Since we observe no significant impact on parameter constraints when fixing $\beta_{\textrm{IA}}$, we decide not to vary $\beta_{\textrm{IA}}$ in our analysis. We also note that there is large uncertainty in the luminosity function at high redshifts. In future work, the effect of different choices for the luminosity function at high redshifts on the IA could be considered.

In summary, our IA model has three free parameters: $A_{\textrm{IA}}$, $\eta_{\textrm{IA}}$  and $\eta^{\textrm{high-z}}_{\textrm{IA}}$. We choose flat priors consistent with \citet{Eifler2020_LSST}, noting that $\eta^{\textrm{high-z}}_{\textrm{IA}}$ is a prior-dominated quantity here.

\subsubsection{Shear calibration}\label{sec:shearcal}
Finally, for the galaxy weak lensing analysis, uncertainties in the shape estimate need to be considered since they are degenerate with the amplitude of the signal. Furthermore, any redshift dependence in these uncertainties can limit our sensitivity to the growth of structure over time \citep{Schaan2017}. A variety of effects such as the inaccuracies in the point spread function, detector effects, and shape selection biases from in-homogeneous galaxy samples all contribute to this shape uncertainty \citep{Massey2013}.

To model the residual shape uncertainty for the weak lensing measurement, we use a multiplicative shear bias factor $m^i$ for each redshift bin $i$
\begin{align}
C_{\gamma \gamma}^{i j}(\ell) & \longrightarrow\left(1+m^{i}\right)\left(1+m^{j}\right) C_{\gamma \gamma}^{i j}(\ell), \\
C_{\delta_{\rm g} \gamma}^{i j}(\ell) & \longrightarrow\left(1+m^{j}\right) C_{\delta_{\rm g} \gamma}^{i j}(\ell).
\end{align}

We follow the assumptions of \citep{Eifler2020_LSST}, using Gaussian priors with a standard deviation of $0.2\%$ for the optimistic and wide scenarios, and $1\%$ standard deviation for the conservative case. The conservative assumption can already be achieved with current generation ground-based experiments like DES \citep{MacCrann2021}. Through an ongoing effort to develop algorithms for shape estimation as well as through the benefit of obtaining higher resolution images from space with Roman, we assume that the achievable uncertainties around launch time will reach approximately our optimistic and wide prior assumptions \citep{Mandelbaum2018,MacCrann2021,Gatti2021,Kannawadi2021}.

Alternatively, \citet{Vallinotto2012,Schaan2017,SimonsObservatory2019} suggested using CMB lensing cross-correlations for self-calibration of the shear bias from data instead of prior calibration. This could both serve as an alternative method if priors can not be achieved as low as projected or as a consistency check of the calibration. We study the self-calibration of shear calibration bias for Roman and SO data in \Cref{res:shearcalib}.

\subsection{Extended cosmological models}\label{sec:extendedcosmologicalmodels}

We have described the calculation of all observables under the assumption of our baseline cosmological scenario described in \Cref{sec:baselinecosmo}. 
Beyond this baseline scenario, we consider a range of other models with modifications to $\Lambda$CDM.

In \Cref{sec:sigma8def}, we consider a model-independent parameterization of the growth of structure, or, equivalently, of the normalization of the linear matter power spectrum, $\sigma_8(z)$. In \Cref{sec:modified_gravity}, we consider modifications to the gravitational potentials. For this, we study a phenomenological model and the Hu-Sawicki $f(R)$ model. For each case, we describe the modifications to our pipeline to include them in our parameter inference. An overview of the free parameters of the extended models is included in \Cref{tab:varied_params}.

\begin{figure*}
\includegraphics[width=17cm]{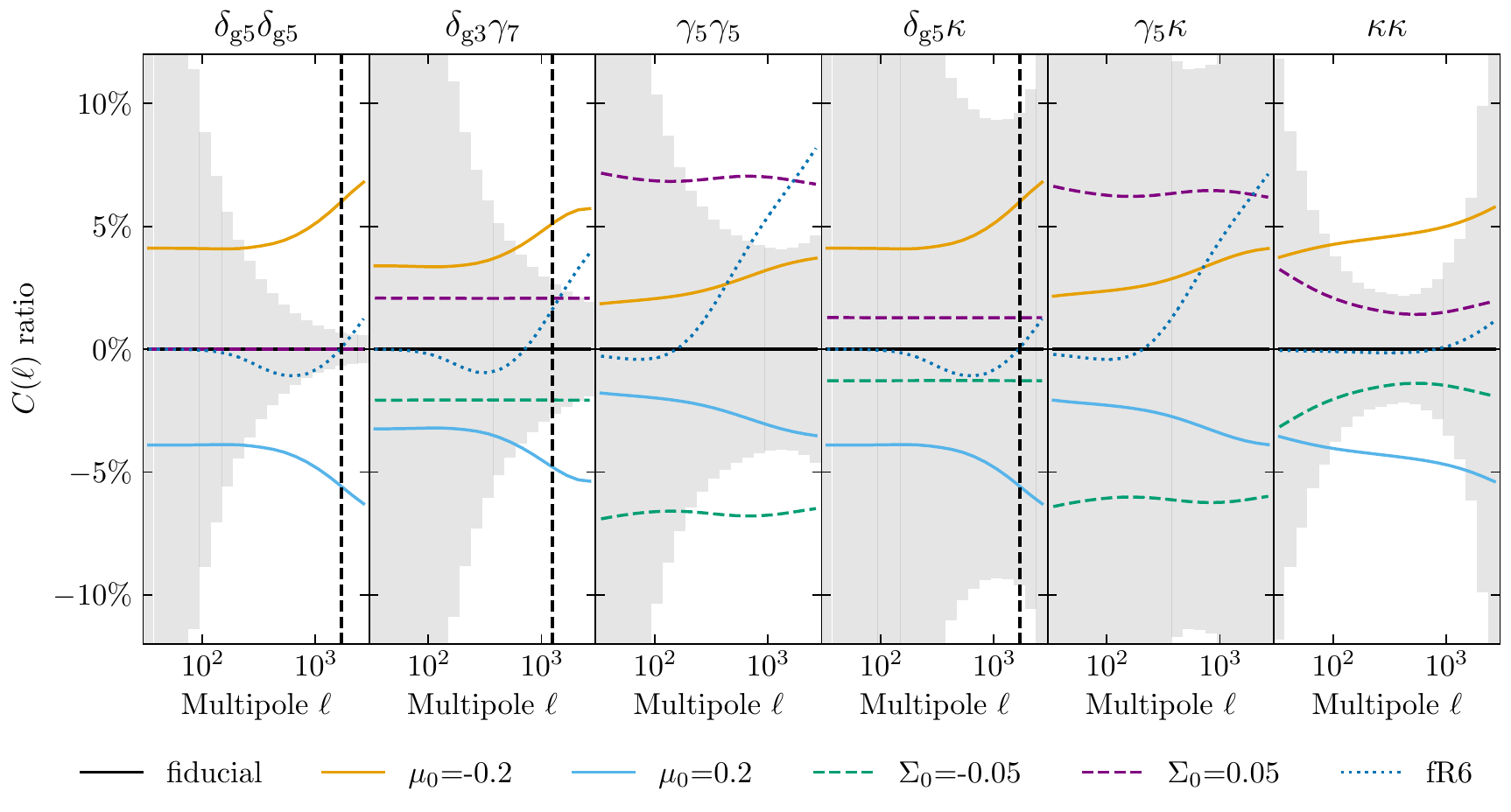}
\caption{The deviation in the observable $C_{AB}(\ell)$ from the fiducial GR for a range of changes in modified gravity parameters. We consider a phenomenological model $(\mu_0,\Sigma_0)$ and Hu-Sawicki f(R) gravity with $f_{R0} = 0$ for GR and $f_{R0} = 10^{-6}$ for the plotted 'fR6'. The specific choices of $A,B \in \{\delta_{gi}$, $\gamma_j$, $\kappa$\} are given above each plot. The modified gravity parameter values are chosen to be approximately consistent with the 1$\sigma$ forecasted constraints from the 6x2pt optimistic HLS analysis. For comparison we show the measurement uncertainty for each of the 20 log spaced bins based on the covariance matrix for the 6x2pt optimistic HLS scenario [gray shaded]. %
For galaxy clustering we use a $\ell$-cutoff scale shown as a black dashed line. }
\label{fig:modgrav}
\end{figure*}

\subsubsection{Model-independent growth of structure ($\sigma_8(z)$)} \label{sec:sigma8def}
 
CMB experiments set tight constraints on the history of growth of structure when assuming the $\Lambda$CDM model \citep{PlanckCollaboration2018}. Galaxy clustering, weak lensing, and CMB lensing can be used to reconstruct the growth history at different times via their sensitivity to the matter power spectrum in different redshift ranges. This offers a powerful test of $\Lambda$CDM by comparing to the extrapolated growth fixed at high redshift by CMB experiments. 
This adds to the discussion around current Stage-III measurements of the growth history that show slight tension between late and early time measurements in the $\Omega_m, \sigma_8$ plane \citep{Joudaki2020,Heymans2020}. 
 
To analyze how well general deviations from the $\Lambda$CDM growth of structure can be constrained we use a model-independent function to quantify the allowed deviation. At the high redshift end ($z\geq 4$) we assume a growth history as in $\Lambda$CDM only allowing for a variation in the normalization, equivalent to varying the normalization of the primordial power spectrum. At the low redshift end ($z<4$) the redshift dependent information from Roman galaxies and its cross correlation allows constraints on deviations in the growth history in a redshift dependent way.

To model this we rescale the linear matter power spectrum by a dimensionless function, $\hat{\sigma}_8^2(z)$, such that
\begin{equation}
    P_{\delta\delta}^{\mathrm{lin}}(k) \rightarrow \hat{\sigma}_8^2(z) P_{\delta\delta}^{\mathrm{lin}}(k),
\end{equation}
and ${\hat{\sigma}_8^2(z)}$ is normalized to unity in $\Lambda$CDM. This makes our prescription equivalent to the approach in \citet{GarciaGarcia2021} where it was applied to current Stage-III data.
To characterize the function we use a cubic spline \citep[using the \textsc{SciPy} implementation,][]{Virtanen2020_scipy} with five anchor points $\hat \sigma_8(z_j)$ at redshifts $z_j = {0,1,2,3,4}$. In order to fully specify the function, we need to provide boundary conditions at both redshift ends. We enforce the first derivative to go to zero at redshift of 4, allowing a smooth transition to high redshifts. 
The value at $z=4$, the last anchor point, effectively sets the high redshift constraint on the power spectrum normalization by affecting the $\Lambda$CDM power spectrum amplitude for $z\geq 4$.
We then simply enforce the second derivative to go to zero at $z=0$, thus allowing for late-time evolution.
Similar approaches of reconstructing the time evolution of cosmological parameters by constraining the anchor points of a spline exist in the literature, e.g. \citet{Vitenti2015,Park2021}.

For this scenario, we fix $\w$ and $\wa$, and also fix the $\sigma_8$ parameter since it is degenerate with multiplying all anchor points by a constant. We use the $\Lambda$CDM growth as the fiducial model and therefore set $\hat \sigma_8(z_j) = 1$ as the fiducial value.
We use the modified linear power spectrum as the basis for the calculation of the non-linear power spectrum, as described in \Cref{sec:angspec}. Note that, by limiting the function to five free parameters, we implicitly assume that any deviation from $\Lambda$CDM is smoother than $\Delta z \approx 1$. Rapidly oscillating deviations would not be captured by our approach. 

\subsubsection{Modified gravity models}\label{sec:modified_gravity} 
In this section, we also consider two modified gravity models.
We start by defining the Friedmann-Robertson-Walker metric in the Newtonian gauge:
\begin{equation}
ds^2 = a^2(\tau) \left[ -(1 + 2\Psi) \mathrm{d}\tau^2 + (1 - 2\Phi) \delta_{ij} \mathrm{d}x_i \mathrm{d}x_j \right] \,,
\end{equation}
where $\Psi$ and $\Phi$ are the scalar perturbations of the metric. 

In GR, the gravitational potential $\Phi$ (the spatial component of the metric perturbations) is related to the matter overdensity in the sub-horizon regime by the Poisson equation:
\be
k^2 \Phi = - 4\pi G a^2 \rho \delta, 
\ee
where $\rho$ is the total matter density, and $\delta$ is the matter density contrast. In the absence of the anisotropic stress, $\Psi = \Phi$ in GR and one often expresses the Poisson equation using $\Psi$. While the equations of motion of \textit{non-relativistic} particles are determined by $\Psi$ (the temporal part of the metric perturbations), the motion of \textit{relativistic} particles is determined by the Weyl potential $\Psi + \Phi$. As a result, the growth of structure is affected by $\Psi$, while the gravitational lensing is affected by $\Psi + \Phi$.

In what follows, we study the deviation from GR by working in the quasi-static approximation and in the absence of anisotropic stress, where the equations governing the evolution of the potentials can be written as \citep{2016ARNPS..66...95J}
\bea
k^2 \Psi = & -  4 \pi G a^2 \mu(k,a) \rho \delta \,, \label{eq:mu} \\
k^2 (\Psi + \Phi) =  & - 8 \pi G a^2 \Sigma(k,a) \rho \delta \,. \label{eq:sigma}
\eea
Here, two quantities parameterize all phenomenology: $\mu$, which parameterizes the effective strength of gravity which impacts the growth of structure; and $\Sigma$, which parameterizes modifications to the gravitational lensing potential. One may also write in the same limits, 
$\Sigma = \frac{1}{2}\mu(1+\gamma)$,
where $\gamma = \Phi/\Psi$ is the slip parameter. In GR, we recover $\mu = 1$ and $\Sigma = 1$.

The two phenomenological parameters $\mu(k,a)$ and $\Sigma(k,a)$ are in principle functions of time and scale. We can map a large class of modified gravity models to $\mu$ and $\Sigma$ in the quasistatic limit \citep{Baker:2014zva, Silvestri:2013ne}. The quasi-static limit is valid when terms in the linearized field equations involving time derivatives of the perturbed quantities are small. For the mapping of parameters in the Horndeski theory (the most general scalar-tensor theory with second-order equations of motion) to $\mu$ and $\Sigma$ space, see e.g. \citet{2020PhRvD.102l3549H} and its source\footnote{\url{https://www.tessabaker.space/images/pdfs/Horndeski_summary.pdf}}.

The observable effects from $\mu$ and $\Sigma$ on the power spectra $C (\ell)$'s are obtained by first solving for the modified growth $D(k,a)$ using~\citep{2012MNRAS.423.3761B}:
\be
\frac{a^2}{D}\dv[2]{D}{a} + \qty(3 + a \dv{\ln E}{a})
\frac{a}{D} \dv{D}{a} = \frac{3}{2}\Omega_{\rm m}(a) \mu,
\ee
where
\be
E(a) = \frac{H(a)}{H_0}, \;\;\;\;
\Omega_{\rm m}(a) = \frac{\Omega_{\rm m,0}\, a^{-3}}{E^2(a)},
\ee
and where $D(k,a)$ and $\mu(k, a)$ can be in general scale-dependent functions. We then obtain the linear matter power spectrum as 
\be
P_{\delta\delta}^{\mathrm{lin}}(k) = P_0(k) [T(k) D(k,a)]^2,
\ee
where $T(k)$ is the transfer function and $P_0(k)$ is the primordial power spectrum as before in our baseline model. Non-linear corrections to the matter power spectrum are added subsequently as described in \Cref{sec:angspec}. 

To obtain $C(\ell)$'s, we compute \Cref{eq:cl} using the modified nonlinear matter power spectrum and the modified lensing kernel, changing \Cref{eq:lensing_kernel_fixed_source} to
\be
W_{\kappa}(\chi, \chi_S) \rightarrow W_{\kappa}(\chi, \chi_S, k=\ell/\chi) = \Sigma(\chi, k=\ell/\chi) W_{\kappa}(\chi, \chi_S).
\ee
These kernels can in principle acquire a scale-dependence as well, in which case we would evaluate $k = \ell/\chi$ using the Limber approximation as is done for the power spectrum $P_{AB}(k = \ell/\chi)$. %

We will consider two modified gravity models in this study. First, a scale-independent phenomenological model $(\mu_0, \Sigma_0)$ where we assert that the phenomenological parameters evolve in time proportional to the dark energy density \citep{DESy1_extendedmodels}. Second, a well-studied scale-dependent model with a chameleon screening mechanism, namely the Hu-Sawicki $f(R)$ model, which exhibits no lensing modifications \citep{Hu2007, 2014MNRAS.440..833A}.

\paragraph*{Phenomenological $(\mu_0,\Sigma_0)$ model:}  \label{sec:mu0sigma0}

As shown above, phenomenological effects of modified gravity models can be described by $\mu$ and $\Sigma$ in the quasi-static limit. So instead of obtaining constraints on individual theory models, one might choose to constrain the evolution of $\mu(k, a)$ and $\Sigma(k, a)$ instead.
Here, however, we start by exploring probes of $\mu(k, a)$ and $\Sigma(k, a)$ in its simplest form and assume no scale dependence. We also impose a time evolution for these parameters that grows proportionally to the dark energy density. In particular, we have a two-parameter model
\begin{equation}
X(a) = 1 + X_0 \frac{\Omega_{\Lambda}(a)}{\Omega_{\Lambda,0}},
\end{equation}
where $X \in \{\mu, \Sigma\}$, and where $\mu_0$ and $\Sigma_0$ are constant. The background expansion history is kept the same as in $\Lambda$CDM. This model was used in a series of work recently \cite[see e.g.][]{2021MNRAS.tmp.2957L, DESy1_extendedmodels, 2019PhRvD..99h3512F}.

This form of time dependence is loosely motivated by the heuristic that if the modified gravity model was to explain the recent phase of acceleration, its phenomenology may become important around the same time that the dark energy density does. This doesn't have to be the case, and is only a speculation, while more complex and better motivated choices are also being explored \citep[e.g.][]{2010PhRvD..81l3508D, 2019LRR....22....1I}.
We choose to adopt this model here not for its theoretical motivation, but rather we view it as a first step toward a full measurement of the functions $\mu(k, a)$ and $\Sigma(k, a)$, which could be measured more fruitfully using, for example, their principal components for a particular survey (see e.g. \citet{2012PhRvD..85d3508H} for LSST; \citet{2013JCAP...08..029A} for a Euclid-like survey). 

In what follows, we choose to forecast constraints on this simplified two-parameter model, in order to study the additional constraining power that CMB lensing may bring. In particular, we will use a figure-of-merit analogous to that used for the $\w\wa$ case:
\begin{equation}
   \textrm{FoM}_{\mu_0 \Sigma_0} \equiv (\textrm{det Cov}(\mu_0, \Sigma_0))^{-1}. \label{eq:FoMmusigma}
\end{equation}

\paragraph*{Hu-Sawicki $f(R)$ model:}

We also consider the Hu-Sawicki $f(R)$ model. In the Jordan frame, the $f(R)$ model takes the form
\be
S = \int \mathrm{d}^4 x\, \sqrt{-g} 
\left[\frac{R + f(R)}{16\pi G} + 
\mathcal{L}_{\mathrm{matter}}
\right],
\ee
where $R$ is the Ricci scalar, $G$ is the gravitational constant, $g$ is the determinant of the metric $g_{\mu \nu}$ and $\mathcal{L}_{\mathrm{matter}}$ is the matter Lagrangian density \citep{DeFelice:2010aj}. One may also cast the theory from Jordan frame where the action for gravity is modified, into the Einstein frame, where the Einstein field equations preserve their GR form, but where there is an additional scalar field non-minimally coupled to gravity. 
Note that for a constant $f$, we recover the cosmological constant.

The Hu-Sawicki $f(R)$ model~\citep{Hu2007}, in particular, is useful because it can reproduce the $\Lambda$CDM background expansion while evading solar system tests by means of the chameleon screening mechanism, in which the scalar field becomes massive and decouples in the regions of high gravitational potential. It is also valued for being free of instabilities and fine-tuning problems unlike some other solutions with a cosmological constant or quintessence field theories.

We adopt the Hu-Sawicki $f(R)$ in the form of a broken power law:
\be
f(R) = -m^2 \frac{c_1 \left(\frac{R}{m^2}\right)^n}{c_2 \left(\frac{R}{m^2}\right)^n + 1},
\ee
where $m^2 = H_0^2 \Omega_{\rm m}$,  and $c_1$, $c_2$ and $n$ are free parameters of the model. By requiring that the model closely mimics the $\Lambda$CDM background expansion history, we obtain
\be
\frac{c_1}{c_2} = 6\frac{\Omega_{\Lambda}}{\Omega_{\rm m}}\;\;\;\;\; \mathrm{and} \;\;\;\; c_2\left( \frac{R}{m^2}\right)^n \gg 1,
\label{eq:hu_sawicki_conditions}
\ee
which eliminates one degree of freedom given cosmological parameters. It is convenient to consider the derivative of $f(R)$
\be
f_R = -n \frac{c_1\left( \frac{R}{m^2}\right)^{n-1}}
{\left[ c_2\left( \frac{R}{m^2} \right)^{n} + 1 \right]^2} 
\approx -n\frac{c_1}{c_2^2} \left( \frac{R}{m^2} \right)^{n+1},
\label{eq:f_R}
\ee
where the approximation uses the second part of \Cref{eq:hu_sawicki_conditions}. The background curvature of the Friedmann-Robertson-Walker metric is given by
\be
\bar{R} = 12 H^2 + 6 \frac{\mathrm{d} H}{\mathrm{d\, ln} a} H = 3m^2 \left(a^{-3} + 4\frac{\Omega_{\Lambda}}{\Omega_{\rm m}} \right),
\ee
and for a flat $\Lambda$CDM expansion history
\be
\bar{R} \approx 3m^2 \left(a^{-3} + 4\frac{\Omega_{\Lambda}}{\Omega_{\rm m}} \right).
\ee
Plugging $\bar{R}$ at its present value (with $a = 1$) into \Cref{eq:f_R}, we obtain the present background value of $f_R$ in terms of $c_1/c_2^2$
\be
{f}_{R0} = -n \frac{c_1}{c_2^2} 
\left( \frac{\Omega_{\rm m}}{3(\Omega_{\rm m}+4\Omega_{\mathrm{\Lambda}})} \right)^{n+1}.
\ee
The model can therefore be fully described by just two parameters: ${f}_{R0}$ and $n$. We recover the GR limit when ${f}_{R0} \rightarrow 0$ or $n \rightarrow \infty$. In what follows, we will fix $n = 1$ and only vary ${f}_{R0}$ in our study of the $f(R)$ model.

To compute the observable effects, we use the mapping onto the phenomenological parameters following ~\citet{2021arXiv210108728L} and~\citet{2009PhRvD..79h3513Z}:
\be
\mu(a, k) = \frac{1+ \frac{4}{3} \frac{k^2} 
{a^2 m^2}}{1+ \frac{k^2} {a^2 m^2}}\;, \;\;\;\;\;
\gamma(a, k) = \frac{1+ \frac{2}{3} \frac{k^2} 
{a^2 m^2}}{1+ \frac{4}{3} \frac{k^2} {a^2 m^2}} \; ,
\ee
where
\be
m^2(a) =  \left(\frac{H_0}{c}\right)^2
\frac{1}{(n+1)|f_{R0}|}
\frac{\left( \Omega_{\rm m} a^{-3} + 4 \Omega_{\Lambda} \right)^{n+2}}{\left(\Omega_{\rm m} + 4 \Omega_{\Lambda} \right)^{n+1}}.
\ee

Note that in this model there is  no modification to the lensing kernel as $\Sigma = \frac{1}{2}\mu(1+\gamma) = 1$ \citep[see appendix of][for a detailed derivation of this fact from the Einstein equations]{2014MNRAS.440..833A}. There is however a scale-dependent growth due to $\mu(a,k)$ which enters through the modified matter power spectrum.

In \Cref{fig:modgrav}, we show the effect of the modified gravity models on the observables. For each type of auto- and cross-correlations, one example is shown, including the expected measurement uncertainty for comparison. As expected, $\Sigma_0$ modifies the lensing observables but does not impact the galaxy clustering observables, because it only impacts the lensing kernel. On the other hand, the $\mu_0$ parameter modifies both types of observables at a similar level with the strongest effects being on small scales.

For the $f(R)$ gravity model, increasing the parameter $f_{\rm R0}$ leads to both a scale and redshift dependent boost in the signal, which is strongest on small scales and at low redshifts. Note that our conservative galaxy clustering cutoffs at small scales limit our sensitivity to the effects of $f(R)$ gravity in this regime. %

\section{Likelihood and sampling}\label{sec:likesamp}

\subsection{Covariance matrix} \label{sec:cov}

\begin{figure}
\includegraphics[width=\columnwidth]{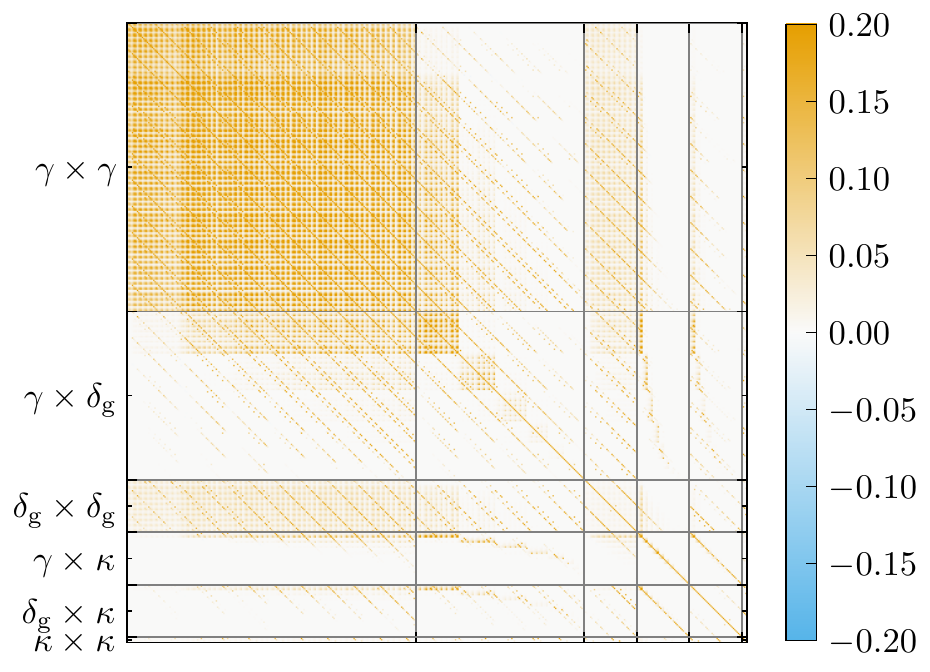}
\caption{The estimated correlation matrix for the optimistic HLS survey+SO case for the full set of observables (6x2pt). The color indicates the correlation with a cutoff at 0.2 to highlight where the contributions from off-diagonals are notable. \label{fig:covariance}}
\end{figure}

We assume that the data vector, obtained by stacking the six types of power spectra for all redshift bin combinations and within scale cuts, follows a multivariate Gaussian likelihood. The mean and fiducial values of this data vector are computed as described in the previous section. The covariance matrix is assumed to be independent of cosmology and is decomposed into a Gaussian and a non-Gaussian contribution.
We estimate the Gaussian part following a standard approach that accounts for the effect of the mask by a simple rescaling given by the observed fraction of the sky, $f_{\rm sky}$, see for example \citet{Zhao2009} and \citet{Pullen2015}. Denoting the elements of the data vector $\tilde{C}_{AB}^{ij}$ as in \Cref{eq:cl}, this is given by
\begin{equation}
    \textbf{C}\left( \tilde{C}_{AB}^{ij}(\ell), \tilde{C}_{CD}^{mn}(\ell^\prime) \right) = \delta_{\ell\ell^\prime} \frac{\left(\tilde{C}_{AC}^{im} (\ell) \tilde{C}_{BD}^{jn}(\ell)+\tilde{C}_{AD}^{in}(\ell) \tilde{C}_{BC}^{jm}(\ell)\right)}{f_{\textrm{sky}}(2 \ell+1) \Delta \ell}\label{eq:covariance}%
\end{equation}
where $\Delta\ell$ is the width of the corresponding bin and 
\begin{align}
\tilde{C}_{AB}^{ij}(\ell) &=C_{AB}^{ij}(\ell)+N_{A}^{i}(\ell) \, \delta_{A B}, \\
N_{\delta_{\rm g} \delta_{\rm g}}^{ij}(\ell) &=\frac{1}{n_{i}^{\textrm{lens}}} \delta_{i j}, \\
N_{\gamma \gamma}^{ij}(\ell) &=\frac{\gamma_{\textrm{rms}}^{2}} {n_{i}^{\textrm{source}}} \delta_{i j}.
\end{align}

The CMB lensing noise power spectrum $N_{\kappa \kappa} (\ell)$ is taken from \citet{SimonsObservatory2019}.
The sky fraction is limited by the footprint of the survey, and we obtain $f_\textrm{sky} \approx 0.048$ and $f_\textrm{sky} \approx 0.436$ for the HLS and wide survey configurations, respectively. For the CMB lensing auto-correlation, however, we can use the full SO footprint, with $f_\textrm{sky}=0.4$. In the wide case, CMB cross-correlations are also limited to $f_\textrm{sky}=0.4$. We note that, for the covariance between maps of different surveys, the factor $f_{\rm sky}$ in \Cref{eq:covariance} is undefined. For this subdominant, off-diagonal part of the covariance, we use the geometric mean of the $f_{\rm sky}$ of the two maps being considered. We note that this is a conservative assumption \citep[for example, in][authors propose using the maximum survey area]{vanUitert:2017ieu}.

The density of galaxies per redshift is given in \Cref{sec:Roman} and \Cref{tab:surveys}, and depends on the Roman survey configuration (HLS or wide).
For the dispersion of galaxy ellipticities, we assume $\gamma_{\textrm{rms}} = 0.37$ (for two components).

\begin{figure*}
\includegraphics[width=\textwidth]{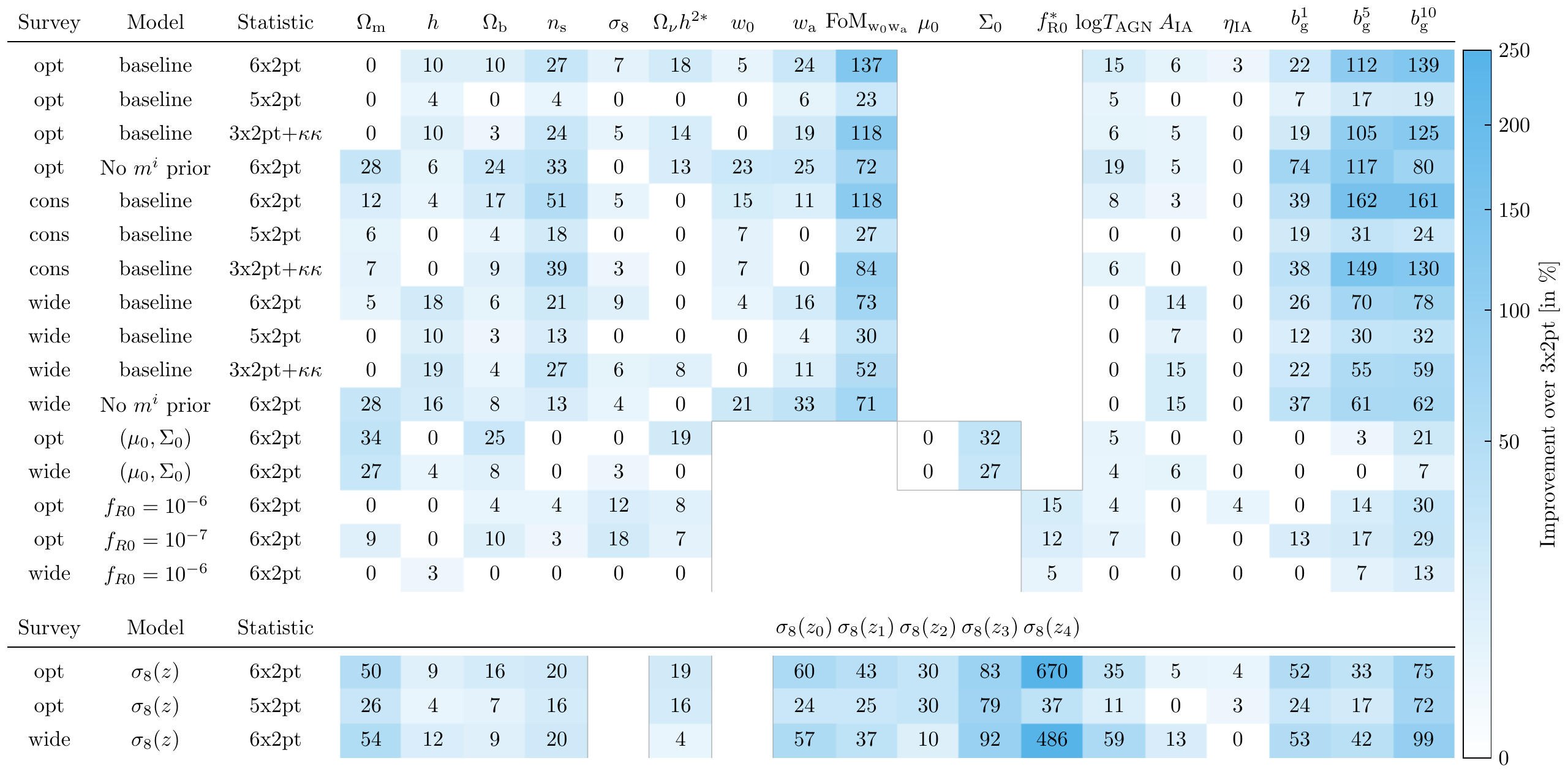}
\caption{Summary table of the full set of likelihood analyses considered showing the improvements for a given survey scenario, cosmological model and statistic when adding SO CMB lensing data relative to the related Roman-only 3x2pt analysis. 6x2pt refers to all correlation functions combined, 5x2pt to only adding the cross-correlations with CMB lensing and $\kappa\kappa$ is the CMB auto-correlation. We show the fractional improvement for the standard deviation of each varied parameter. (*) For all $\Omega_\nu h^2$ and $f_{\textrm{R0}}$ in the $f_{\textrm{R0}} = 10^{-7}$ scenario we instead compare the 95\% upper limit. Stronger improvement is color coded blue. Improvements below 2.5\% are shown as 0.}\label{fig:resultstable} 
\end{figure*}

The non-Gaussian contributions to the covariance matrix are split into the connected four-point component from the trispectrum and the super-sample covariance due to the contribution of super-survey modes. We use the \cosmolike software to compute these contributions \citep{Eifler2014cosmolike,Krause2017cosmolike,Krause2020cosmolike}, extending the covariance matrix used in \citet{Eifler2020_HLS}, following a procedure similar to \citet{Eifler2020_LSST,Fang2021}.

The covariance matrix is recomputed for each survey configuration, as they assume different redshift distributions (see \Cref{figure:bins}), galaxy densities, and survey area. We show the covariance matrix for the optimistic HLS scenario in \Cref{fig:covariance}.

\subsection{Sampling} \label{sec:sampling}
 
As detailed in \Cref{sec:baselinecosmo} and \Cref{tab:varied_params}, the parameter space of our baseline model comprises 54 varied parameters, some of which are expected to be unconstrained and/or limited by the prior boundaries (for instance, the neutrino density $\Omega_\nu h^2$ or some nuisance parameters).
Purely employing a Fisher forecast approach would, therefore, yield biased results. Sampling this parameter space with standard Markov Chain Monte-Carlo (MCMC) or nested sampling method is computationally expensive, however, since parts of the pipeline (in particular \hmcode) require a few seconds per parameter point. We, therefore, devised a fast sampling approach that allows us to test several models and survey configurations. For each configuration, we proceed as follows:
\begin{enumerate}
    \item We start by computing the Fisher matrix of our likelihood model, which requires computing derivatives of the data vector with respect to the model parameters. This is achieved with a finite difference method. %
    By inverting the Fisher matrix $\mathbf{F}$, we obtain a first approximation of the posterior, given by a multivariate Gaussian distribution, centered on our fiducial parameter values and with covariance $\mathbf{F}^{-1}$.
    \item We then build an approximate but differentiable
    model of the theoretical data vector which can be rapidly and efficiently sampled with a Hamiltonian Monte Carlo approach. To do so, we first draw points from the Fisher approximation of the posterior and run our full pipeline on these points. We then apply MOPED compression \citep{Heavens2000} to the data vector using the pre-computed derivatives and covariance matrix. This reduces the dimension of the data vector to that of the parameter space and its covariance becomes the identity matrix, by construction. 
    We then use the \bayesfast\footnote{\url{https://github.com/HerculesJack/bayesfast}} package, introduced in \citet{Jia2019_bayesfast}, to fit a polynomial (quadratic) model of the compressed data vector as a function of model parameters and sample this model with a Hamiltonian Monte Carlo method (HMC) \citep{Hoffman2011}.
    We use 14 independent chains of 10,000 samples, discard the first 1,000 as warm-up, and use those points as input to repeat this entire procedure. The second set of chains, for which we verify convergence with the standard Gelman-Rubin statistic $R$ \citep{Gelman1992_GelmanRubinStatistic}, provides a second approximation to the posterior distribution.
    \item As the last step, we use (truncated) importance sampling \citep{Ionides2008_truncatedimportancesampling} for the full posterior model with the uncompressed data vector, to remove any potential bias stemming from the polynomial model approximation and the compression step. We use the \bayesfast HMC chains as input, which we thin by a factor of 10 to ensure independence, giving a total of 12,600 posterior samples in our final results. The effective number of weighted samples is typically a few thousand: e.g.\ 3,349 for our baseline optimistic 6x2pt result. We systematically check that the constraints, before and after importance sampling the points, are consistent with each other, to ensure that our samples are a good representation of the posterior distribution.
\end{enumerate}

This approach allows us to sample the posterior distribution in a few hours on a single node of 14 physical cores. We tested it against \polychord \citep{Handley2015polychord_1of2,Handley2015polychord_2of2}, an established nested sampling tool that has been used for cosmological parameter inference (see \Cref{sec:code_valdiation}). Our approach saves an order of magnitude in computing time compared to \polychord, while giving consistent results. We note that it only works for forecasts where the fiducial values are known, but unlike a Fisher matrix approach, we need not assume Gaussianity for the inferred constraints.

We use GetDist \citep{Lewis2019getdist} to handle our chains, including enforcing flat prior boundaries and kernel density estimation smoothing of marginalized constraint plots.

\section{Results}\label{sec:results}

In this section, we present the results of our forecasts. The presented constraints are based on the inference scheme presented in \Cref{sec:sampling} and use the set of free parameters listed in \Cref{tab:varied_params}.

We report on the expected improvements when adding CMB lensing from SO to Roman galaxy clustering and weak lensing data. We first explore the case of our baseline scenario described in \Cref{res:baseline}. Then, in \Cref{res:pess}, we vary the survey strategies to consider how improvements change in a conservative HLS survey scenario and, in \Cref{results:wide_survey}, a proposed wider survey that trades photometric depth with a larger covered sky area. In \Cref{res:shearcalib}, we explore how well the multiplicative shear uncertainty can be self-calibrated from data when removing assumed prior calibration. Finally, in \Cref{res:extmod}, we explore constraints on a range of extended cosmologies.

We consider various combinations of the statistics in our analysis. We refer to the combination of LSS tracers only as "3x2pt", comprised of ($\gamma\gamma, \delta_{\rm g} \delta_{\rm g} , \gamma \delta_{\rm g}$). When adding both auto- and cross-correlations with the CMB lensing we use the term ``6x2pt", and include ($\gamma\gamma, \delta_{\rm g} \delta_{\rm g} , \gamma \delta_{\rm g} , \gamma \kappa, \delta_{\rm g}  \kappa, \kappa \kappa$).  We further differentiate the CMB lensing improvements by considering constraints when adding only the cross-correlations of CMB lensing ($\gamma\gamma, \delta_{\rm g} \delta_{\rm g} , \gamma \delta_{\rm g} , \gamma \kappa, \delta_{\rm g}  \kappa$), referred to ``5x2pt", and by only adding the CMB lensing auto-correlation, ``3x2pt+$\kappa \kappa$".

In \Cref{fig:resultstable} we provide an overview for results showing the main parameter improvements for our different scenarios. 
The improvement between 3x2pt and runs including CMB lensing data is given as the percentage improvement on the standard deviation constraint, defined as $\sigma_{\textrm{p}} (\textrm{3x2pt})/{\sigma_{\textrm{p}}(\textrm{6x2pt})}- 1$. In the overview table, we show improvements above a nominal cutoff of 2.5\%, but below we focus our discussion on where this technique shows significant improvements in parameter constraints and point out where improvements are only marginal.

\subsection{Optimistic Roman HLS survey scenario}
\label{res:baseline}

We first consider our baseline scenario of an optimistic HLS survey from Roman with dark energy parametrized using $\w \wa$ and including massive neutrinos. To determine the impact of including CMB lensing data from SO, we compare the prospective constraints.

\begin{figure}
\includegraphics[width=\columnwidth]{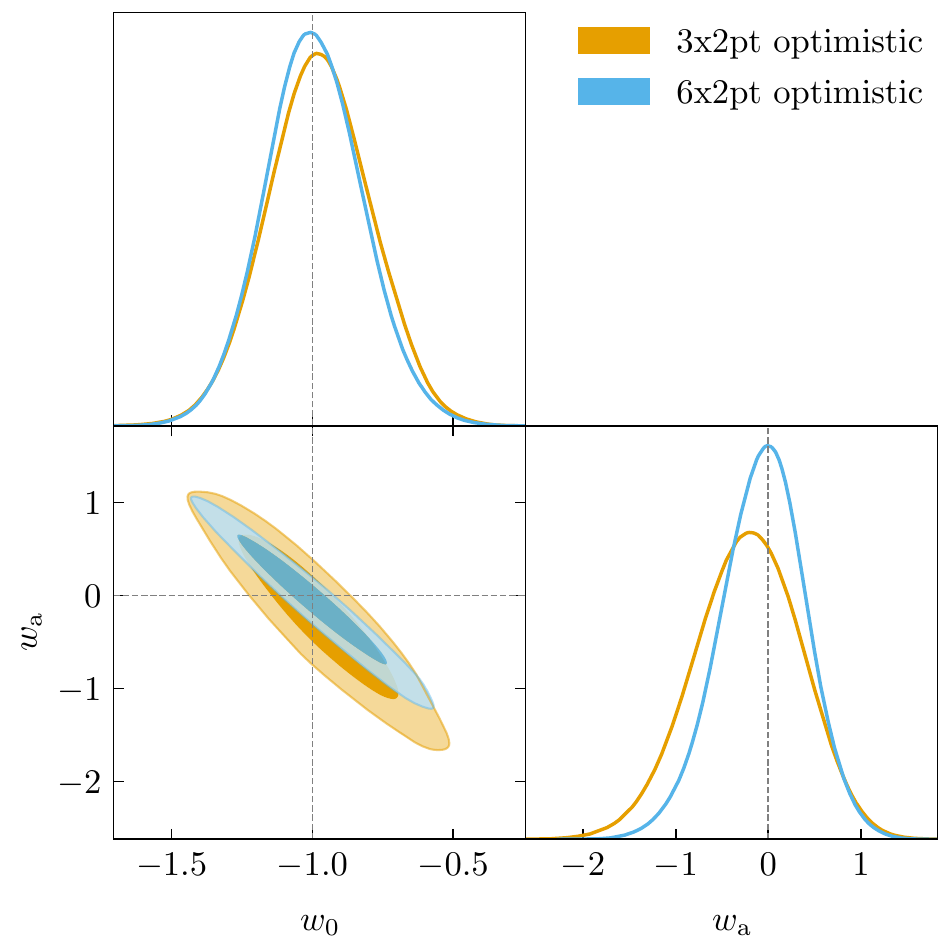}
\caption{Marginalized constraints on $w_{\textrm{0}}$ and $w_{\textrm{a}}$ for the optimistic HLS survey scenario. Shown are the 3x2pt [orange] and 6x2pt [blue] constraints. The addition of SO CMB lensing improves the FoM by a factor 2.4. \label{fig:darkenergy}}
\end{figure}

As shown in \Cref{fig:resultstable}, for the baseline scenario, comparing 6x2pt and 3x2pt, we find that constraints on all the core parameters except $\Omega_{\rm m}$ are improved by the addition of CMB lensing. The most pronounced effects are  in  $n_{\rm s}$ and $\wa$ with 27\%, 24\% and 18\% improvements respectively. 
The improvements in $\w$ and $\wa$ translate into a factor of 2.4 improvement in the dark energy FoM for dark energy, defined in  \Cref{eq:FoM}, as shown in \Cref{fig:darkenergy}.

\begin{figure}
\includegraphics[width=\columnwidth]{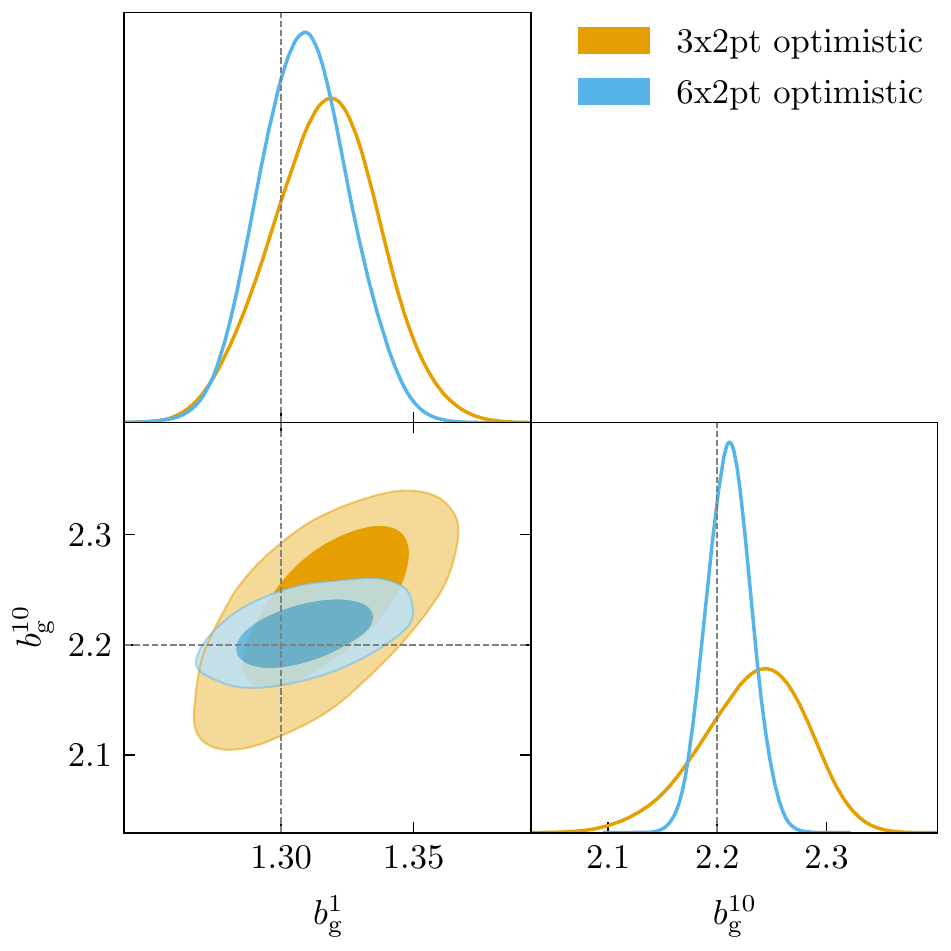}
\caption{Marginalized constraints on the galaxy bias parameters in the lowest ($b_g^1$) and the highest redshift bin ($b_g^{10}$) for the 3x2pt [orange] and 6x2pt [blue] observables in the optimistic HLS survey scenario.}
\label{fig:galaxybias}
\end{figure}

Galaxy bias parameter constraints are significantly improved by the addition of CMB lensing, increasing as one moves to the higher redshift bins, from 22\% at low redshift to 140\% at high redshift. %
Due to the lack of background sources at the high redshift end of our galaxy samples, the uncertainty of the galaxy bias increases with redshift. CMB lensing acts as an additional background source that is also unaffected by galaxy bias. It thereby predominantly improves the more poorly constrained higher redshift bins. In \Cref{fig:galaxybias} we show an example of the improvement on the galaxy clustering biases constraints, for the lowest and highest redshift bins. The inclusion of CMB lensing correlations, in the 6x2pt case, reduces the correlation between the different galaxy clustering bias bins. This also reduces the degeneracies in the 6x2pt between cosmological and galaxy bias parameters, relative to the 3x2pt. %

For the optimistic HLS survey, the improvements are dominated by the CMB lensing auto-correlation: The FoM for dark energy improves by a factor of 1.2 for the 5x2pt but by a factor of 2.4 for 3x2pt+$\kappa \kappa$ compared to the baseline 3x2pt. 
Similarly for the other cosmological parameters and the galaxy bias, most of the improvement is driven by the CMB lensing auto-correlation. 

This can be understood in terms of the sky fraction the respective statistics cover. The CMB lensing auto-correlation uses data with $f_{\textrm{sky}} = 0.4$, while the cross-correlations only cover the limited area of the HLS, which is \SI{2000}{\deg \squared} or $f_{\textrm{sky}} \approx 0.048$. 
This difference changes for the wide Roman survey scenario, discussed in \Cref{results:wide_survey}.

\subsection{Conservative Roman HLS survey scenario}\label{res:pess}

In addition to the optimistic scenario, we also consider a conservative version of the HLS survey in which we assume weaker priors on the shear calibration and photometric redshift nuisance parameters, as well as a larger photometric redshift uncertainty as summarized in \Cref{tab:varied_params}. 
Relaxing the prior assumptions reduces the achievable cosmological constraints. As an example, the dark energy FoM for the 3x2pt analysis in the conservative scenario is a factor of 2.6 smaller than that for the optimistic one. The addition of auto- and cross-correlations with CMB lensing from SO helps to compensate for this. The conservative 6x2pt achieves a FoM for dark energy of 21 which is comparable to the FoM of 23 forecasted for the optimistic 3x2pt scenario. In \Cref{tab:FOMde} we summarize the forecasted FoM for all scenarios.

\begin{table}
\begin{center}
\begin{tabular}{lllll}
\toprule
Dark Energy FoM & 3x2pt & 5x2pt & 3x2pt+$\kappa \kappa$ & 6x2pt \\ \hline
Optimistic HLS                             & 23  & 29  & 51 & 55    \\
Conservative HLS                                   & 9.5   & 12  & 17  & 21    \\
Wide                                         & 140 & 190 & 220  & 250 \\ %
\toprule
\end{tabular}
\caption{The dark energy Figure of Merit (FoM) for the three different Roman survey scenarios: optimistic HLS, conservative HLS and wide. The FoM for each scenario is calculated for the Roman LSS data alone (3x2pt) and thefull Roman and SO CMB lensing data (6x2pt). Combinations of the LSS data with just the CMB auto-correlation (3x2pt+$\kappa\kappa$) and just the cross-correlations (5x2pt) are also considered. Results rounded to 2 significant figures. \label{tab:FOMde}}
\end{center}
\end{table}

The shear calibration and photometric redshift nuisance parameters remain mostly prior dominated, as they also were in the optimistic case. Only for the highest redshift bin, the photometric redshift parameters show marginal improvements of 6-16\% which are not significantly affecting overall cosmological constraints.

There are greater improvements, relative to the 3x2pt, in the galaxy bias parameters than for the optimistic case: from $\sim$40\% at low redshift to $\sim$160\% at the high redshift. As with the optimistic case, the principal driver of the improvement is the auto-correlations (the 3x2pt+$\kappa\kappa$ is markedly better than the 3x2pt alone), however, the cross-correlations (5x2pt) also make a non-negligible contribution. 

\subsection{Wide Roman survey scenario}\label{results:wide_survey}

In addition to optimistic and conservative HLS scenarios, that would survey \SI{2000}{\deg \squared}, we also consider a wide survey scenario where Roman would perform a shallower survey over a much larger area, \SI{18000}{\deg \squared}
(see \Cref{tab:surveys}).

Overall the dark energy FoM for the wide survey is markedly higher than the optimistic HLS survey with the 3x2pt wide scenario FoM 6.2 times greater than the one for the optimistic HLS scenario. This is because the wide scenario has lower shot noise by observing a greater total number of galaxies and lower cosmic variance due to the larger sky coverage.
We find that the addition of CMB lensing significantly helps tighten the dark energy constraints for the wide scenario, though at a slightly lower level than for the HLS survey scenarios, with the 6x2pt FoM a factor of 1.7 greater than that from the 3x2pt.  The combination of the wide survey and CMB lensing data projects a dark energy FoM that is more than an order of magnitude (11 times) greater than the FoM of the optimistic HLS survey alone, without CMB lensing data (3x2pt). 
This strengthens the science case for both a wide survey during a possible extended mission of the Roman Space Telescope and the joint analysis with CMB data.

In the wide survey scenario, the improvement in galaxy bias parameters is less than that for the HLS survey scenario but still significant: 26\% at low redshift to 80\% at high redshift. Shear calibration and photometric redshift errors are again prior limited. 

We again differentiate the improvements from adding only the CMB lensing cross-correlations (5x2pt) and only adding the CMB lensing auto-correlation (3x2pt+$\kappa\kappa$). The FoM for dark energy for the %
for the 5x2pt and 3x2pt+$\kappa\kappa$ are respectively 1.3 and 1.5 times that of the 3x2pt.
Consistently for the other cosmological parameters, about one third of the improvement comes from the cross-correlation with CMB lensing and about two thirds come from the CMB lensing auto-correlation. Therefore, in the wide scenario with more overlap with SO data, the cross-correlations are more relevant than for the optimistic HLS survey scenario.

\subsection{Self-calibration of shear calibration bias}\label{res:shearcalib}

\begin{figure}
\includegraphics[width=\columnwidth]{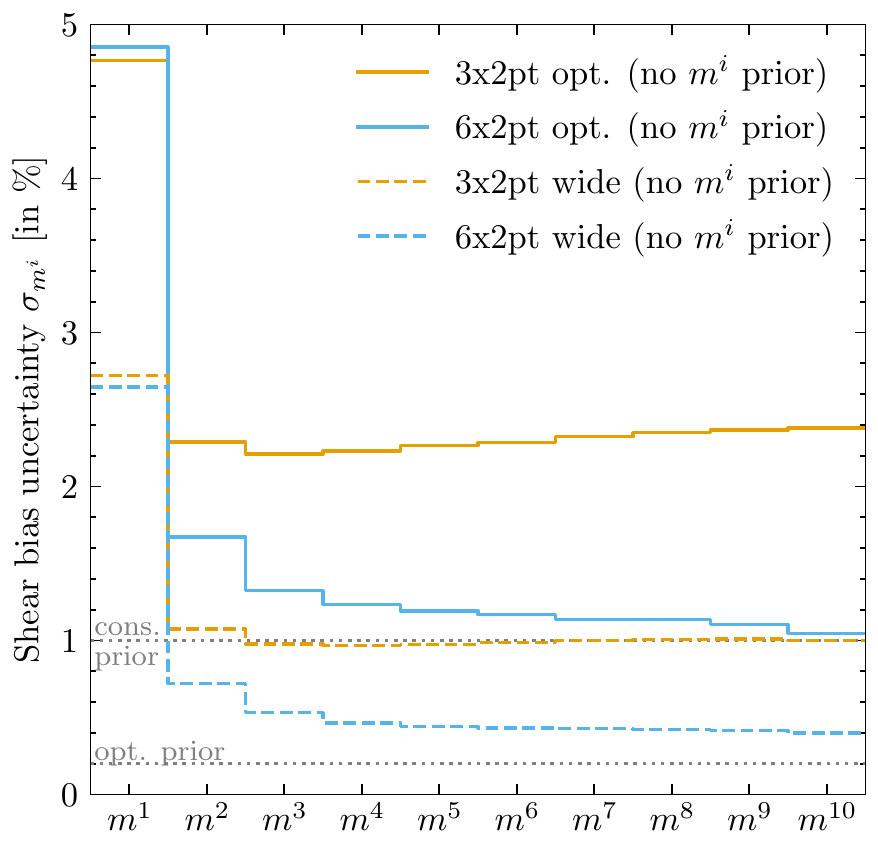}
\caption{The effect of Roman survey self-calibration [3x2pt, orange] and CMB lensing [6x2pt, blue] on Roman shear calibration bias,  in the optimistic HLS scenario [full lines] and wide survey scenario [dashed lines], %
if the Roman priors on $\textrm{m}^i$ [gray lines] are removed. In both cases adding CMB lensing information significantly improves self-calibration, especially at high redshifts.}
\label{fig:shearmagnification}
\end{figure}

\begin{figure}
\includegraphics[width=\columnwidth]{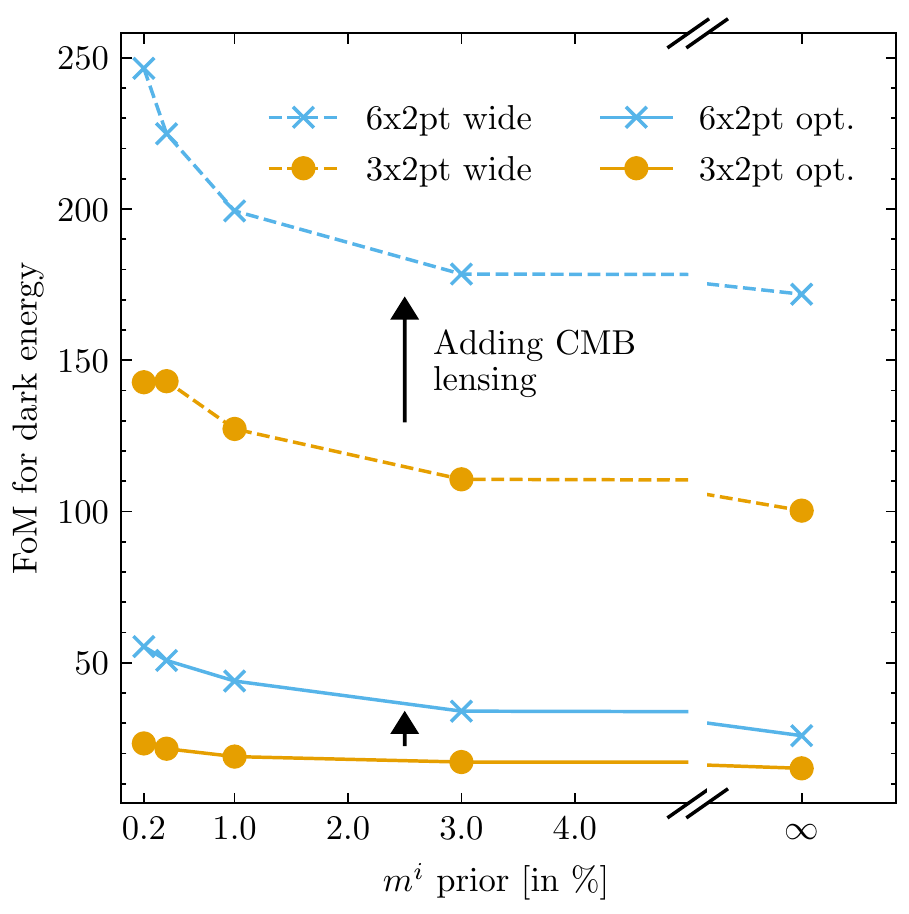}
\caption{The forecasted dark energy FoM as the assumed prior on the multiplicative shear bias $m^i$ is varied. Shown are 6x2pt [blue] and 3x2pt [orange] constraints for both the optimistic HLS [full lines] and wide survey [dashed lines] scenarios. }
\label{fig:shearbias_vs_FoM}
\end{figure}

In the previous sections, we employed strong priors on the shear calibration nuisance parameters. As discussed in \Cref{sec:angspec}, uncertainty in the shear calibration is sourced by a range of instrumental and observational effects. If the projected performance levels are not attained, then the calibration from data alone, through the addition of CMB lensing, can provide a complementary approach. %
The extra information contained in the CMB lensing cross-correlations provides a complementary way to extract out the cosmological lensing signal of the foreground structure with distinct background sources. We expand on previous work by \citet{Vallinotto2012,Schaan2017,SimonsObservatory2019} on this self-calibration, which was mainly focused on LSST.

To determine the level to which the two lensing signals can be used in this way, we rerun our baseline cosmology with the Gaussian priors on $m^i$ removed, using a conservative sampling range of $[-0.1, 0.1]$ for all $m^i$. 
\Cref{fig:shearmagnification} shows the achievable uncertainties for the 6x2pt and the 3x2pt. The self-calibrated constraints on $m^i$ for the optimistic HLS and wide survey scenarios are shown as solid and dashed lines respectively, with 3x2pt in orange and 6x2pt in blue. In both cases, we see a strong improvement in the constraint when including CMB lensing, that increases with redshift. For the optimistic HLS survey scenario, the standard deviation constraint improves by 40\% in the second bin, and the improvement increases to 130\% for the last bin. In absolute numbers, $\sigma_{m^{10}}$ reaches approximately 1\% for the highest redshift bin which is equivalent to the conservative prior. For the wide survey scenario, $\sigma_{m^i}$ improves by 50\% in the second bin, and the improvement increases to 150\% for the last bin. In absolute numbers, $\sigma_{m^{10}}$ reaches 0.4\% for the highest redshift bin, which is factor 2 higher than our wide prior of 0.2\%. Therefore, self-calibration of the shear calibration bias at high redshift is more effective for a wide survey than the deep HLS survey. Adding CMB lensing information from SO to the HLS survey can be used to self-calibrate the shear bias at the level of the conservative prior at high redshift.

Beyond stating the level of self-calibration we want to study the effect on the achievable constraints on cosmology. For this, we present a range of runs in \Cref{fig:shearbias_vs_FoM}, where we show the FoM for dark energy as we step-wise increase the prior on the shear bias, from none to the optimistic assumption of a 0.2\% prior. From this, we can see that when including CMB lensing without any prior information on $m^i$ we can already achieve a larger FoM for dark energy than a 3x2pt with a 0.2\% prior. This is the case in both an optimistic HLS and wide scenario. Furthermore, when having both prior information on $m^i$ and CMB lensing data the FoM for dark energy increases further. In fact, the difference between 6x2pt and 3x2pt gets larger as the assumed prior gets tighter for both optimistic HLS and wide scenarios. To summarize, CMB lensing information can effectively be used to compensate weaker priors on $m^i$ than anticipated, but, to achieve the strongest constraint on cosmology, both a strong prior calibration of the shear bias and of CMB lensing data are ideal.

\subsection{Constraints on extended cosmological models}\label{res:extmod}

\subsubsection{Model-independent growth constraints: $\sigma_8(z)$}\label{res:sig8}

\begin{figure*}
\includegraphics[width=\textwidth]{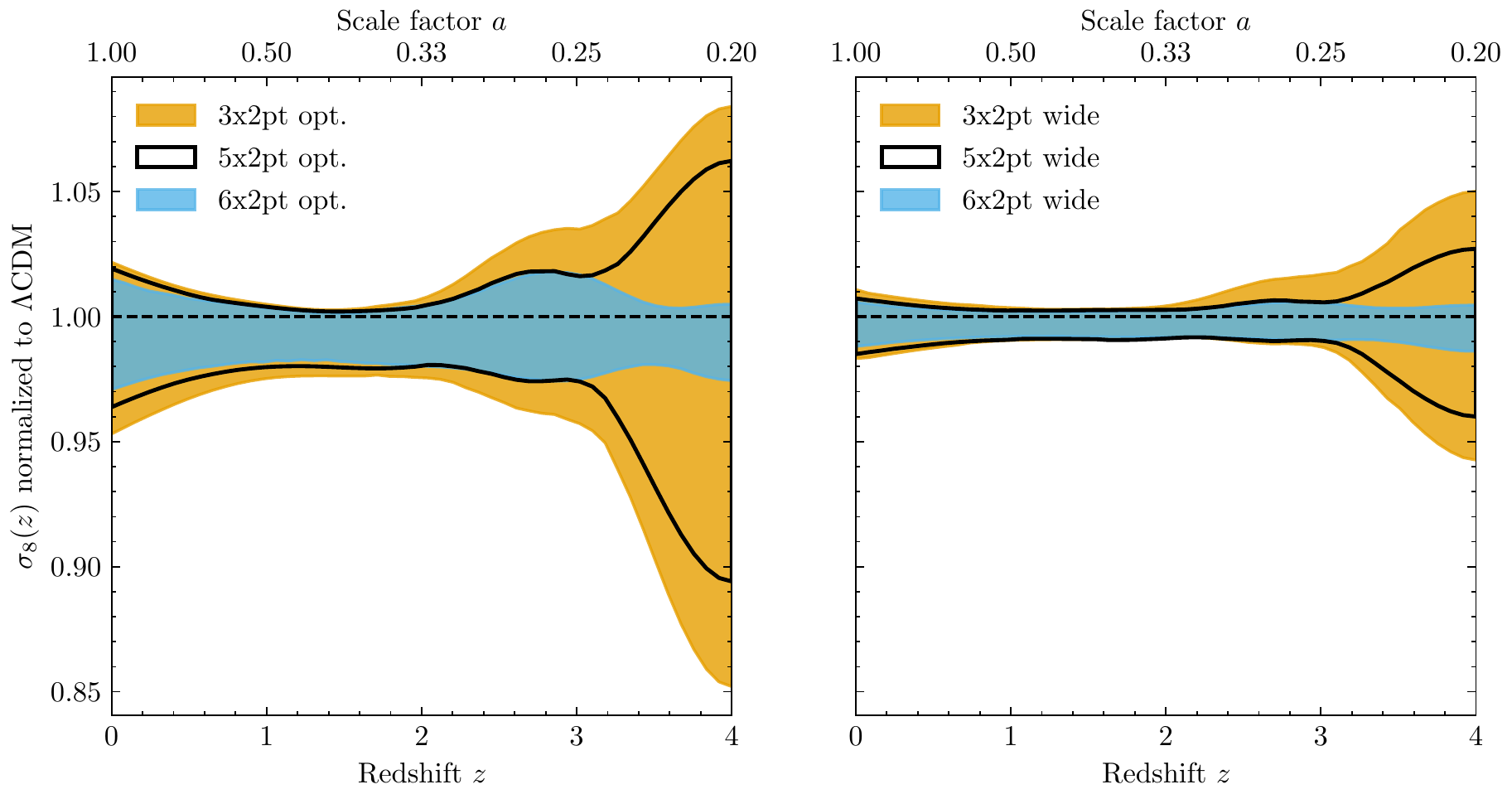}
\caption{Constraints on $\sigma_8(z)$ relative to $\Lambda$CDM for the optimistic HLS survey scenario [left] and wide survey scenario [right]. The bands, interpolated between five free parameters at z=0,1,2,3 and 4, show the forecasted 68\% constraints  achievable
with only Roman data [3x2pt, orange], when including only CMB lensing cross-correlations [5x2pt, black thick line] and including all CMB lensing data [6x2pt, blue].  }
\label{fig:sigma8ofz}
\end{figure*}

Next, we study, for the first time, Roman's capability to constrain the growth of structure in a model-independent way, by measuring the redshift dependence of $\sigma_8(z)$. To do so, we use a spline 
with 5 anchor points $\hat{\sigma}_8(z_j)$ as described in \Cref{sec:sigma8def}.
In \Cref{fig:sigma8ofz} we show the 1$\sigma$ constraint on our parametrized $\sigma_8(z)$ for the optimistic HLS scenario on the left, and the wide survey scenario on the right. We show the 3x2pt constraint band in orange and the 6x2pt constraint in blue. We also show the case of only adding the CMB lensing cross-correlations to Roman data (5x2pt) as a black line. Overall, we find that Roman 3x2pt can constrain $\sigma_8(z)$ to about 2\% in the redshift range $0.7-2$ in the optimistic scenario. This improves to about 1\% in the wide survey scenario, thus providing a powerful test of the growth of structure as predicted by $\Lambda$CDM.

There is a strong improvement when adding the CMB lensing information. The improvement is across all redshifts, with the biggest difference at low and high redshifts. The 5x2pt constraint shows that a large fraction of the improvement at lower redshift comes from the cross-correlations between CMB lensing and Roman data. In contrast, at the high redshift end, adding the CMB lensing auto-correlation leads to significant further improvement. Overall, the constraint on $\sigma_8(z)$ from the wide survey scenarios is stronger at all redshifts. In both cases, we find that adding CMB lensing data from SO to Roman 3x2pt data leads to significantly improved constraints, especially at high redshift: 
$\hat \sigma_8 (z_5 = 4)$ improves by about a factor 8 for the optimistic HLS scenario and a factor 6 for the wide scenario. Overall, we find that Roman 6x2pt can constrain $\sigma_8(z)$ to within 3\% and 2\% across the redshift range from 0 to 4 in the optimistic HLS and wide scenarios respectively.

Our constraints on the galaxy bias parameters limit the constraints we can achieve on $\sigma_8(z)$. The addition of CMB lensing data allows better constraints on the galaxy bias, which translates to better constraints on $\sigma_8(z)$\footnote{Knowing the galaxy bias exactly would give an order of magnitude better constraints on $\sigma_8(z)$ for the optimistic HLS scenario.}. 
We also note that our highest redshift anchor point $\hat \sigma_8 (z_5 = 4)$ is at the high redshift end of Roman clustering and weak lensing data while the CMB lensing kernel is sensitive even to redshifts beyond 4 (see \Cref{eq:lensing_kernel_fixed_source}). Therefore, a strong improvement for this parameter when adding CMB lensing information is consistent.

\subsubsection{Modified Gravity Model I: $(\mu_0,\Sigma_0)$}\label{res:muSig}

\begin{figure}
\includegraphics[width=\columnwidth]{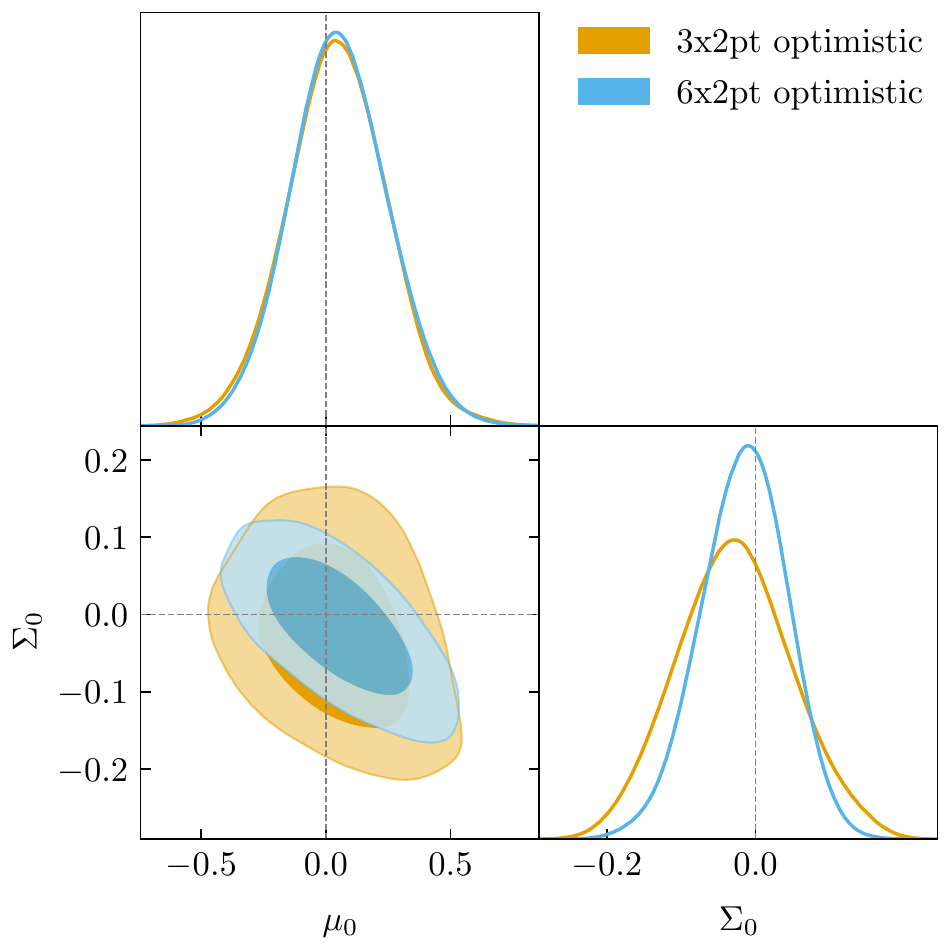}
\caption{Marginalized constraints on $\mu_0$ and $\Sigma_0$ modified gravity model parameters for the 3x2pt [orange] and 6x2pt [blue] data for the optimistic HLS survey scenario. The addition of CMB lensing data improves the ${\rm FoM}_{\mu_0 \Sigma_0}$ by 64\%.}
\label{fig:musigma}
\end{figure}

We now consider our first modified gravity model parameterized by the phenomenological parameters $\mu_0$ and $\Sigma_0$, as described in \Cref{sec:mu0sigma0}. Recall that $\mu_0$ parameterizes the deviation from the strength of gravity in GR, and $\Sigma_0$, deviations from the lensing kernel in GR. As a result, we find that the addition of CMB lensing mainly helps to constrain the $\Sigma_0$ parameter that describes the modification to lensing. In \Cref{fig:musigma} we show the marginalized constraint for these two parameters for the optimistic HLS scenario. As seen in the 1D posteriors, the constraints on $\mu_0$ barely changed while the error in $\Sigma_0$ tightened by 32\% when CMB lensing is added. More specifically, our 3x2pt analysis forecasts $\sigma_{\Sigma_0} = 0.077$ and $\sigma_{\mu_0} = 0.20$, whereas for the 6x2pt analysis $\sigma_{\mu_0}$ remains the same and $\sigma_{\Sigma_0} = 0.059$.
Moreover, we can calculate the figure-of-merit for the $(\mu_0,\Sigma_0)$ parameter combination as defined in \Cref{eq:FoMmusigma}. The addition of CMB lensing going from 3x2pt to 6x2pt improves $\textrm{FoM}_{\mu_0\Sigma_0}$ by a factor of 1.6 in the optimistic HLS scenario.

\begin{table}
\begin{center}
\begin{tabular}{lllll}
\toprule
$\textrm{FoM}_{\mu_0 \Sigma_0}$ & 3x2pt & 6x2pt \\ \hline
Optimistic Roman HLS scenario                           &  69 & 110   \\
Wide Roman scenario                                         & 380 & 540 \\ %
\toprule
\end{tabular}
\caption{Figure-of-merit for the parameter combination ($\mu_0,\Sigma_0$) in the phenomenological modified gravity model. The FoM improves by a factor of 1.6 (1.4) when the SO CMB lensing data (6x2pt) is added to the Roman galaxy clustering and weak lensing data (3x2pt) in the optimistic (wide) scenario. Results are rounded to 2 significant figures. \label{tab:FOMmusig}}
\end{center}
\end{table}

We also considered the improvement for the wide survey scenario. Here we find qualitatively the same in that the $\mu_0$ constraint changed negligibly and that there is a substantial improvement of 27\% in the constraint for $\Sigma_0$. In particular, the projected 1D constraints in the wide case for a 3x2pt analysis are $\sigma_{\Sigma_0} = 0.039$ and $\sigma_{\mu_0} = 0.07$; for the 6x2pt analysis the former improves to $\Sigma_0 = 0.031$. Moreover, the $\textrm{FoM}_{\mu_0 \Sigma_0}$ for 6x2pt is 1.4 times the one for 3x2pt. We summarize the $\textrm{FoM}_{\mu_0 \Sigma_0}$ in \Cref{tab:FOMmusig}. Overall, the $\textrm{FoM}_{\mu_0 \Sigma_0}$ for a wide survey scenario is higher than for the optimistic reference HLS survey scenario by a factor of 4.9. This shows that the wide survey will be much more constraining on the $(\mu_0,\Sigma_0)$ modified gravity model when combined with CMB lensing.

\subsubsection{Modified Gravity Model II: Hu-Sawicki $f(R)$}\label{res:fR}

\begin{figure}
\includegraphics[width=0.92\columnwidth]{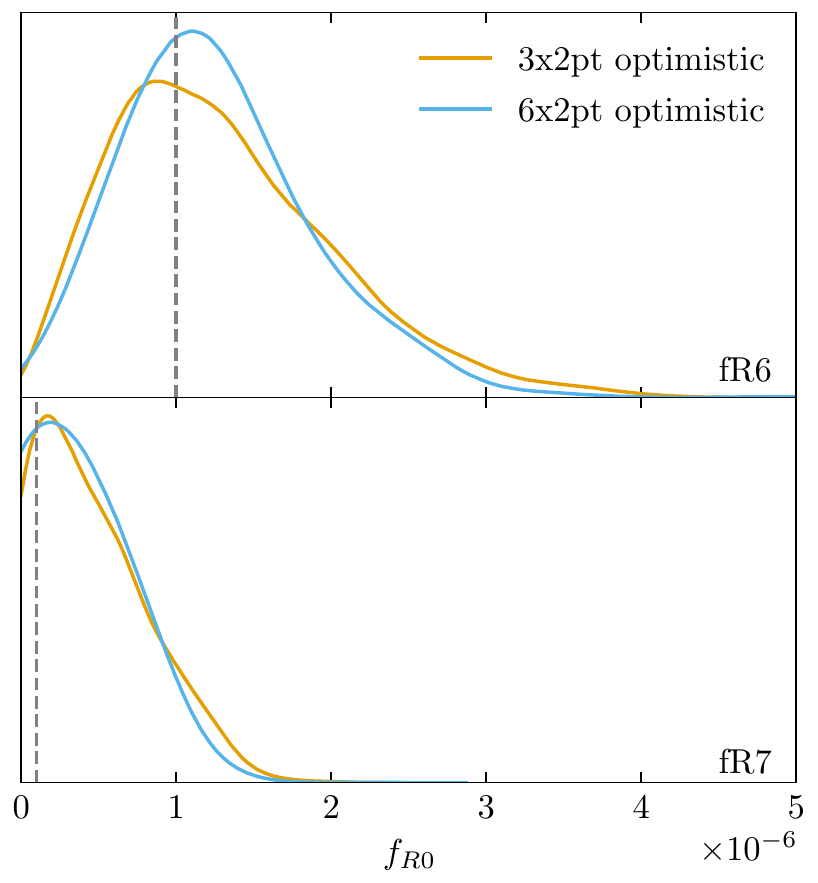}
\caption{Marginalized constraints on f(R) gravity for two fiducial values of $f_{\rm R0}$, referred to as fR6 [upper] and fR7 [lower]. Shown are the 3x2pt [orange] and 6x2pt [blue] constraints for the optimistic HLS survey scenario.\label{fig:fR}}
\end{figure}

Finally, we also consider a Hu-Sawicki $f(R)$ model of modified gravity as described in \Cref{sec:extendedcosmologicalmodels}. In this model, the background expansion is preserved as in $\Lambda$CDM and there is no modification to the lensing kernel. However, there is a scale-dependent modification to the growth of structures which affects our observables as shown in \Cref{fig:modgrav}. %
Assuming fixed $n=1$, for fiducial values of $f_{\rm R0} \gtrsim 10^{-6}$ deviations start in the linear regime and are possibly measurable by probes of the matter power spectrum \citep{Hu2007}. For fiducial values of $f_{\rm R0} \lesssim 10^{-7}$ deviations move into the non-linear regime, making them difficult to distinguish from astrophysical effects.

We present forecasts for fiducial values of $f_{\rm R0} = 10^{-6}$ and $f_{\rm R0} = 10^{-7}$ (hereafter "fR6" and "fR7"), exploring how well the former could be detected, and what upper limits we could place on the latter. In \Cref{fig:fR} we show the 1D marginalized constraints on $f_{R0}$ for a 3x2pt and 6x2pt analysis under the assumption of the optimistic HLS scenario. The addition of CMB lensing does not provide a significant improvement: For fR6 the parameter $f_{R0}$ improves by 15\%, and for fR7 the 95\% upper limit improves by 12\%. We attribute this only marginal improvement to the fact that the $f(R)$ model does not modify the lensing kernel and mainly modifies the low redshift growth of structure while the lensing kernel of CMB lensing only peaks at $z\sim 2$. 
Assuming the optimistic HLS scenario, the full 6x2pt analysis for fR6 yields %
${f_{\rm R0} = (1.28^{+0.45}_{-0.75}) \cdot 10^{-6}}$, where the deviation from GR can be detected at 1.7$\sigma$ in this parameter. For fR7, the 95\% lower limit achieved is $f_{\rm R0} < 1.1 \cdot 10^{-6}$. %

We also run our forecast for fR6 assuming our wide scenario. 
Here, the improvement from adding CMB lensing information is not significant (only 5\% difference for $f_{\rm R0}$). Under our assumptions for the survey, the full 6x2pt analysis would achieve %
${f_{\rm R0} = (1.03^{+0.20}_{-0.24}) \cdot 10^{-6}}$,
which is a 4.3$\sigma$ detection of deviations from GR. %

\section{Discussion}\label{sec:discussion}

In this work, we consider combining weak lensing and clustering data from the Roman Space Telescope with CMB lensing data from the Simons Observatory (SO). For our modeling, we include the important relevant astrophysical effects and calibration uncertainties. These include neutrinos, baryonic effects on dark matter clustering, as well as uncertainties in the galaxy bias, in the intrinsic alignment, in the shear calibration, and in the photometric redshift estimation.  
We consider different survey scenarios for Roman: a deep HLS survey covering \SI{2000}{\deg \squared} in an optimistic and conservative scenario, and also a shallower but wider survey scenario covering \SI{18000}{\deg \squared}. %
This lets us explore the trade-off between depth and area, and follows the current proposed plans for Roman. %
We explored how CMB lensing can improve constraints on a range of extended cosmological models with a focus on where significant improvements can be achieved due to SO CMB lensing. Here we summarize the main results.

We find that adding CMB lensing data from SO to Roman weak lensing and clustering data approximately doubles the FoM for dark energy. The exact improvement varies between a factor of 1.7 and 2.4, depending on the survey configuration. The improvement is strongest for the deeper HLS survey scenario, where most of the improvement is driven by the much larger area CMB lensing auto-correlation. For a wide survey scenario with more overlap with SO CMB lensing data, the cross-correlations between the datasets ($\gamma \kappa, \delta_{\rm g} \kappa$) become more relevant, contributing about a third of the improvement, whereas the CMB lensing auto-correlation contributes about two thirds. %

Among our set of nuisance parameters, we find that CMB lensing information improves the constraints on the galaxy biases significantly, whereas the baryonic model and IA parameters show only marginal improvements. Beyond the scope of this work, more careful modeling of contributions like non-linear and post-Born effects on the cross-correlations with CMB lensing will require additional modeling to analyze observational data but we expect similar overall conclusions. 
Given our assumptions for the achievable calibrations of the Roman survey, there is no improvement on the photometric redshift errors and the shear calibration from adding CMB lensing. They remain prior-dominated at the level of their calibration. However, under a scenario where the calibration of shear biases is weaker than anticipated, or as a consistency check of the calibration, CMB lensing provides a powerful alternative way to calibrate the shape uncertainties from data alone. 

In particular, we refine statements about the achievable self-calibration of shear bias for Roman+SO in this work. Consistent with previous work by \citet{Schaan2017,SimonsObservatory2019}, we find that the CMB lensing significantly improves self-calibration, especially at high redshift. More precisely, the shear calibration in the highest redshift bins for the optimistic HLS survey can be self-calibrated to about 1\% by using SO CMB lensing data, similar to the level of our conservative prior. For a wide survey scenario, the fractional improvement by adding CMB lensing data is similar and the absolute calibration reaches even lower values of 0.4\%. 
For both the optimistic HLS and wide scenario the FoM for dark energy achievable when adding CMB lensing but assuming no shear prior is better than no CMB lensing information but an optimistic shear bias prior. 
To maximize cosmological constraints, both a strong prior on the shear bias and CMB lensing information are ideal.  

Furthermore, Roman has the potential to constrain other models beyond $\Lambda$CDM. We explored the capabilities of Roman in combination with CMB lensing from SO to constrain a range of interesting extensions. For this, we maintain our extensive set of nuisance parameters to report a forecast with a realistic error budget. 

First, we considered the constraints on a model-independent growth of structure as a function of redshift, $\sigma_8(z)$. Assuming that the deviations from $\Lambda$CDM are smooth within redshift intervals of 1, a 6x2pt analysis can constrain the deviations in the growth of structure up to $z \sim 4$ at the 3\% level in an optimistic HLS scenario (see \Cref{fig:sigma8ofz}). The forecasted constraints are even better for the wide survey scenario, about 1\%. 
In both scenarios, the addition of CMB lensing data plays a crucial role and is especially helpful at the high redshift end.

We also considered modifications to the gravitational potentials based on two examples: a phenomenological model $(\mu_0,\Sigma_0)$ and the Hu-Sawicki $f(R)$ model. 
For our two modified gravity models, we find that adding CMB lensing information principally tightens constraints on the parameters directly affecting the lensing kernel, i.e. $\Sigma_0$. In contrast, the $\mu_0$ parameter in the $(\mu_0,\Sigma_0)$ model, and the $f_{R0}$ parameter of the $f(R)$ model, are not significantly improved by adding CMB lensing, as they only modify the growth of structure, whose constraints are driven by the Roman HLS data.
Overall, the Roman and SO data in combination provide a powerful probe able to achieve a constraint of $\sigma_{\Sigma_0} = 0.059$ and $\sigma_{\mu_0} = 0.20$ for the $(\Sigma_0, \mu_0)$ model, and a 95\% upper limit of $f_{\rm R0} < 1.1 \times 10^{-6}$ for the $f(R)$ model in the optimistic HLS scenario. This shows how Roman+SO can go significantly beyond current Stage-III constraints from weak lensing and clustering data \citep{DESy1_extendedmodels}.
For the wide survey scenario, constraints get even tighter, reaching $\sigma_{\Sigma_0} = 0.031$ and $\sigma_{\mu_0} = 0.07$; for the $f(R)$ model, a deviation from GR at the level of $f_{\rm R0} = 10^{-6}$ would be detected with high significance (4.3$\sigma$). 

In ascertaining the potential of combining the Roman galaxy clustering and weak lensing data with CMB lensing from SO, we have taken care to include a set of nuisance parameters %
representing a realistic account of the total error budget. We have also considered different possible Roman survey configurations. %
The intent is to inform survey selection considerations, and the potential interplay with concurrent CMB surveys, as Roman moves towards launch in the second half of the 2020s. It also allows and motivates 
comparisons with other cosmological tracers that will be analyzed from the Roman HLS, such as clusters, SNe and voids \citep{Eifler2020_HLS}, as well as SO and CMB-S4 probes like the primary CMB, thermal and kinematic Sunyaev-Zeldovich effects, and the cosmic infrared background \citep{SimonsObservatory2019}.

We conclude that there is great potential in a combined analysis of Roman galaxy clustering and weak lensing data with CMB lensing data from SO.
This is one of many synergies that will be possible with Stage-IV photometric, spectroscopic, and CMB datasets that are coming online in the current decade covering significant overlapping fractions of the sky for the first time. Combinations of these tracers will put tight constraints on dark energy and new physics.

\section*{Acknowledgments}

We thank Agnès Ferté for providing the code for solving the modified growth equation. We thank Nicholas Battaglia for helpful discussions about CMB lensing and SO. We thank the other members of the \textit{Cosmology with the High Latitude Survey} Roman Science Investigation Team for helpful discussions. 

The work of Lukas Wenzl and Rachel Bean is supported by NASA grant 15-WFIRST15-0008 \textit{Cosmology with the High Latitude Survey} Roman Science Investigation Team, NASA ATP grant 80NSSC18K0695, NASA ROSES grant 12-EUCLID12-0004 and  DoE grant DE-SC0011838. CD and BJ were supported in part by the US Department of Energy grant DE-SC0007901. XF is supported by the Department of Energy grant DE-SC0020215, the NASA ROSES ATP 16-ATP16-0084 grant, and the BCCP Fellowship. Part of this work was done at the Jet Propulsion Laboratory, California Institute of Technology, under a contract with the National Aeronautics and Space Administration. Government sponsorship acknowledged.

Resources supporting this work were provided by the NASA High-End Computing (HEC) Program through the NASA Advanced Supercomputing (NAS) Division at Ames Research Center.

Calculations in this paper use High Performance Computing (HPC) resources supported by the University of Arizona TRIF, UITS, and RDI and maintained by the UA Research Technologies department.

\section*{Data Availability}

\cosmosis \citep{Zuntz2015} is available under \url{https://bitbucket.org/joezuntz/cosmosis}. \hmcode \citet{Mead2021} is available under \url{https://github.com/alexander-mead/HMcode}. \cosmolike \citep{Eifler2014cosmolike,Krause2017cosmolike,Krause2020cosmolike} is shared under reasonable request to the corresponding authors of those papers. The main components of the code underlying this work are available as a repository under  \url{https://github.com/WFIRST-HLS-Cosmology/RomanxCMB}. Any other data underlying this article will be shared on reasonable request to the corresponding author.

\bibliographystyle{mnras}
\bibliography{master.bib,additional_references.bib}

\appendix
\section{Code validation}\label{sec:code_valdiation}

We performed a range of tests on the code for our forecast to ensure sufficient accuracy. 
First we consider the pipeline to calculate observables based on cosmological and nuisance parameters as described in \Cref{sec:formalism}. We compare the full data vector consisting of all observable power spectra (hereafter \cosmosis) with the largely independent pipeline used in \citet{Eifler2020_HLS} (hereafter \cosmolike). Note that we use the \cosmolike pipeline for the calculation of the non-linear part of the covariance for this work as well. 
\cosmolike does not implement our IA model and used \halofit without modeling of baryonic effects. So for the comparison, we use our \cosmosis pipeline with \halofit and IA turned off. %
Overall agreement between the two implementations is good: half of all datapoints points agree within 1.7\% and 90\% of points agree within 4.4\%. This is consistent with our expectations: We use the Transfer fitting functions from \citet{Eisenstein1999} which claim an accuracy of about 5\%. This is mainly due to neglecting baryonic oscillations which are not relevant for our forecast. Overall this shows that our implemented pipeline has sufficient accuracy.

\begin{figure*}
\includegraphics[width=18cm]{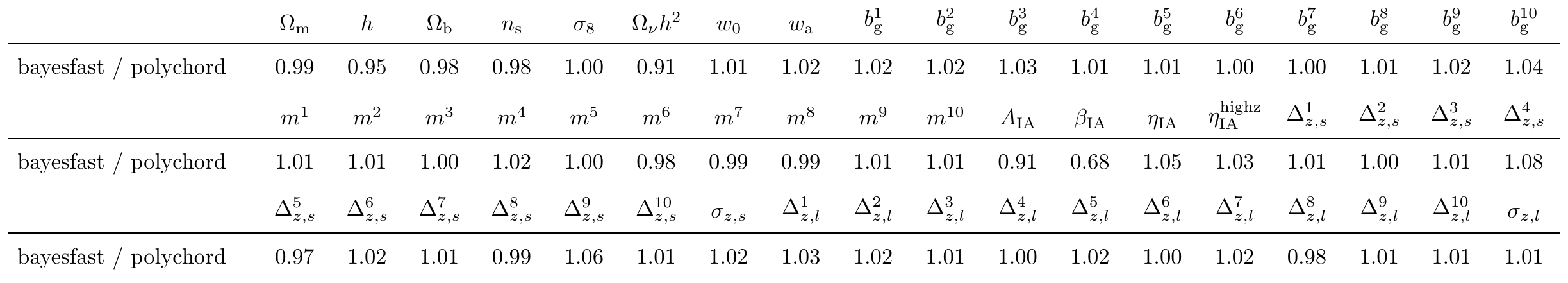}
\caption{Validating accuracy of our parameter inference code labeled \bayesfast by comparing the inferred parameter variances to the established \polychord. Shown are the ratios of the variances for each parameter for the case of optimistic HLS scenario in our baseline cosmology. Based on this we decided to fix $\beta$, see \Cref{sec:angspec}. Values within a few percent of 1 indicate excellent agreement. %
\label{fig:codevalidation_polychord}}
\end{figure*}

To test our parameter inference code described in \Cref{sec:sampling} (hereafter \bayesfast) we compare with the well-established nested sampling method \polychord \citep{Handley2015polychord_1of2,Handley2015polychord_2of2}. To make this computationally feasible, we use a simplified pipeline using \halofit to calculate the non-linear power spectrum, instead of \hmcode, speeding up likelihood calculations by one order of magnitude. With this setup, we therefore use all parameters from \Cref{tab:varied_params}, except $\log T_{\rm{AGN}}$. %
For the \polychord run, we use 800 live points, number of repeats of 60, a tolerance of 0.1, and spend 40\% of time in the subspace of fast parameters ($b_{\rm g}^{i}, m^{i}$). Other \polychord parameters are left at their default values.

\Cref{fig:codevalidation_polychord} shows the comparison between our parameter inference approach and \polychord. Shown are the ratios of the variances of 1D marginalized constraints. The overall agreement is very good, with only a few percent difference between the two methods in general. We note that \bayesfast struggles to explore the tails of the asymmetric distribution for the sum of neutrino masses ($\Omega_\nu h^2$, 9\% difference in the standard deviation). It also struggles to explore the non-Gaussian degeneracy between $A_{\rm{IA}}$ and $\beta_{\rm{IA}}$, which leads to a 32 \% lower standard deviation in $\beta_{\rm{IA}}$. We decide to fix $\beta_{\rm{IA}}$ which, we have confirmed, does not have a significant effect on overall constraints since its uncertainty is considered through the degenerate $A_{\rm{IA}}$ parameter.
The agreement between \polychord and \bayesfast is excellent on cosmological parameters of interest.
We conclude that our parameter inference approach achieves sufficient accuracy while being two orders of magnitude faster than the \polychord approach we compared to.

\label{lastpage}
\end{document}